\documentclass[conference]{IEEEtran}

\usepackage{cite}
\usepackage{amsmath,amssymb,amsfonts}
\usepackage{algorithmic}
\usepackage{url}
\usepackage{graphicx}
\usepackage{textcomp}
\usepackage{pgfplots}
\usepackage{xcolor}
\usepackage[colorlinks=true,
            linkcolor=blue,
            urlcolor=blue,
            citecolor=blue]{hyperref}
\usepackage{subfigure}
\usepackage{hyperref}
\usepackage{comment}
\usepackage{subcaption}
\usepackage{booktabs}

\begin{document}

\title{An Information-Theoretic Perspective\\ on LLM Tokenizers}
\author{%
\IEEEauthorblockN{Mete Erdogan$^{1}$$^{*}$, Abhiram Gorle$^{1}$$^{*}$, Shubham Chandak$^{2}$, Mert Pilanci$^{1}$, Tsachy Weissman$^{1}$}
\IEEEauthorblockA{$^{1}$Department of Electrical Engineering, Stanford University, USA, $^{2}$AWS Annapurna ML}
\{merdogan, abhiramg, pilanci, tsachy\}@stanford.edu, shubhamchandak94@gmail.com}

\maketitle
\renewcommand{\thefootnote}{\fnsymbol{footnote}}
\setcounter{footnote}{0}
\footnotetext[1]{Equal contribution, listed alphabetically.}
\renewcommand{\thefootnote}{\arabic{footnote}}

\begin{abstract}

Large language model (LLM) tokenizers act as structured compressors: by mapping text to discrete token sequences, they determine token count (and thus compute and context usage) and the statistical structure seen by downstream models. Despite their central role in LLM pipelines, the link between tokenization, compression efficiency and induced structure is not well understood. We empirically demonstrate that tokenizer training scale \textit{redistributes} entropy: as training data grows, the token stream becomes more diverse in aggregate (higher unigram entropy) yet markedly more predictable in-context (lower higher-order conditional entropies), indicating that tokenization absorbs substantial short-range regularity although these gains degrade under train-test domain mismatch. To ground these observations, we first benchmark i) pretrained GPT-family tokenizers as black-box compressors across various domains, and ii) learned tokenizers across configurations spanning vocabulary size, training scale, and domain. Next, we study tokenization as a transform for universal compression and introduce a compression-aware BPE variant. Finally, we adopt a channel lens and introduce capacity-utilization metrics to analyze tokenizer behaviour and outline implications for downstream modeling. Put together, our results expose various trade-offs between compression, induced structure, and robustness under domain shift, and motivate principled, compression-aware tokenizer design.

\looseness=-2

\end{abstract}

\section{Introduction}
\label{intro}
\noindent
Tokenization is a foundational yet underexplored component of modern large language model (LLM) pipelines. By mapping raw text (bytes or characters) into sequences over a finite vocabulary, a tokenizer induces a discrete representation whose sequence length, redundancy, and statistical structure directly affect computational cost and downstream modeling~\cite{sennrich2016neural, radford2019gpt2}. Tokenization serves two interlinked objectives: it segments text by defining a codebook over variable-length substrings, and compresses byte or character sequences into a finite-alphabet symbol stream. This induces an inherent trade-off between segmentation and compression. Classical source coding schemes can achieve rates arbitrarily close to entropy, but their bit-level representations often obscure the local structure that learning systems exploit~\cite{lz77, rissanen1984universal, goose2024compressing}. Conversely, tokenizers optimized solely for modeling performance may produce redundant representations, increasing sequence length and computational cost~\cite{tokenizercontrol2024}. Effective tokenization must balance compression efficiency, preservation of linguistic structure, and suitability for predictive modeling.

Recent work has begun to probe trade-offs in tokenizer design, showing that minimizing token count alone cannot improve downstream performance, and mismatches between tokenizer training and deployment domains can degrade
performance~\cite{tokenizercompression2024, tokenizeropt2024}. Yet, a comprehensive characterization of how tokenizer design impacts compression, induced statistical structure, and robustness remains a largely open problem.

\textbf{Our Contributions:} In this work, a) we first benchmark pretrained GPT-family tokenizers as black-box compressors across multiple domains, comparing their behaviour to classical compressors such as gzip, zstd \cite{lz77, collet2018zstandard}, b) we then conduct a controlled study of learned tokenizers across varying training data, domains and vocabulary sizes to find a consistent scaling trend: as tokenizer training data increases, unigram entropy rises while higher-order conditional entropies decrease, consistent with classical $n$-gram modeling results~\cite{chen1996, towardTokenizationTheory2024}, c) next, we study tokenizers under train--test domain mismatch, necessitating domain-aware tokenizer training and evaluation, d) we also examine the interaction with universal compressors, and find that tokenization can act as a useful preprocessing transform and e) complementing these results, we finally present a simple channel lens to summarize how vocabulary size and token-frequency skew govern ``capacity utilization'', that can guide principled, compression-aware tokenizer design.

\section{Related Work}
\label{related}
Prior work highlights a trade-off in LLM tokenization between compression efficiency and the learnability of the induced token representation. \cite{goose2024compressing} shows that while arithmetic coding achieves near-optimal compression, its bitstreams lack linguistic structure and can severely degrade model performance without additional constraints, underscoring that \emph{compression alone need not yield learnable representations}. Likewise, \cite{tokenizercompression2024} finds that minimizing token count is an unreliable proxy for downstream quality, since vocabulary coverage and segmentation boundaries materially affect performance. In contrast, \cite{tokenizeropt2024} shows that when tokenization is aligned with the data distribution, more compressive tokenizers can improve perplexity and generalization. Together, these results suggest tokenizers must jointly balance rate reduction, linguistic granularity, and domain fit. \looseness=-2

Some recent approaches propose principled frameworks drawing on rate--distortion theory and the information bottleneck to construct representations that discard predictable redundancy while preserving
task-relevant or semantic information~\cite{tishby2000information, shani2025tokens, young2025radio}. Building on these insights, our work aims to provide a unified evaluation of compression, entropy, and downstream performance across domains, with specific attention to how \emph{tokenizer training sample length} affects robustness under distribution shift.

\section{Compressibility of Pretrained LLM Tokenizers}
First, we analyze the compressibility properties of four pretrained GPT-family tokenizers: \texttt{gpt2} \cite{radford2019gpt2}, \texttt{p50k\_base} \cite{brown2020gpt3}, \texttt{cl100k\_base} \cite{achiam2023gpt}, \texttt{o200k\_base} \cite{tiktoken} \footnote{the numbers in each tokenizer denote the tokenizer vocabulary size}, encompassing OpenAI’s tokenizers from GPT-2 to GPT-5. Treating each tokenizer as a black-box compressor, we ask: \emph{where do GPT tokenizers shine as compressors, and where are they brittle?}

The tokenizers are evaluated across three domains: news, code and math. For each domain, we use standard Hugging Face (HF) datasets: C4 \cite{raffel2020exploring} for news, CodeParrot \cite{CodeParrot} for code, and GSM8K \cite{cobbe2021gsm8k} for math. On each corpus, we evaluate on $10k$ samples (yielding upto 1M characters per domain), first measure the intrinsic compressibility of the raw UTF-8 bytes using \textbf{zstd} (bits per character), and then tokenize the same text with each baseline to compute tokens per character. This yields the compression–tokenization tradeoff plot shown below in Fig.~\ref{fig:scatter}a). Appendix~\ref{app-A} briefly outlines the correlation between tokenization and compression in the above setup.
\vspace{-0.4cm}
\begin{figure}[ht]
    \centering

    \subfigure[Compressibility across various domains]{
        \includegraphics[width=0.7\linewidth]{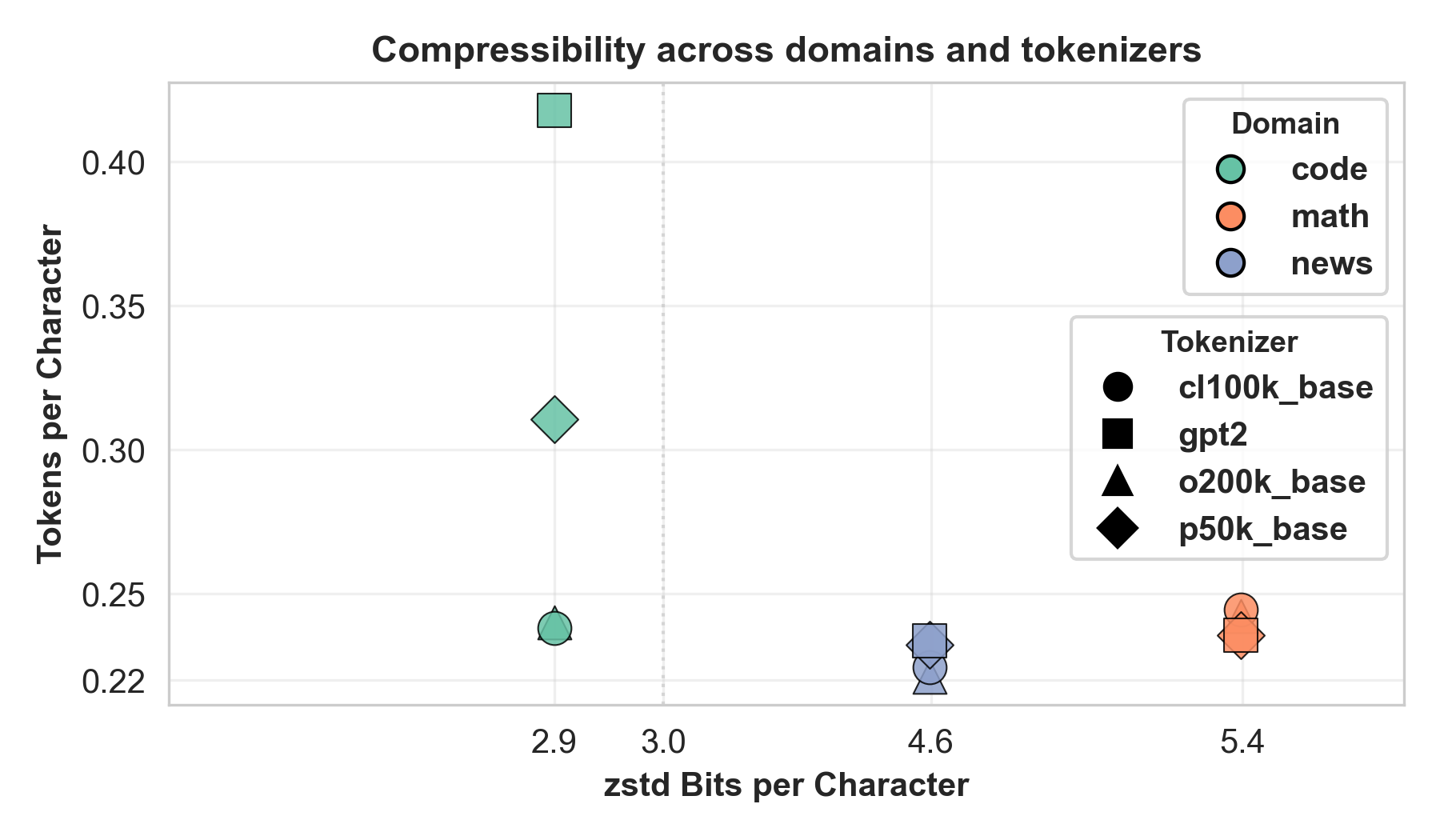}
    } \hspace{-0.45cm}
    \subfigure[Cross-lingual robustness]{
        \includegraphics[width=0.85\linewidth]{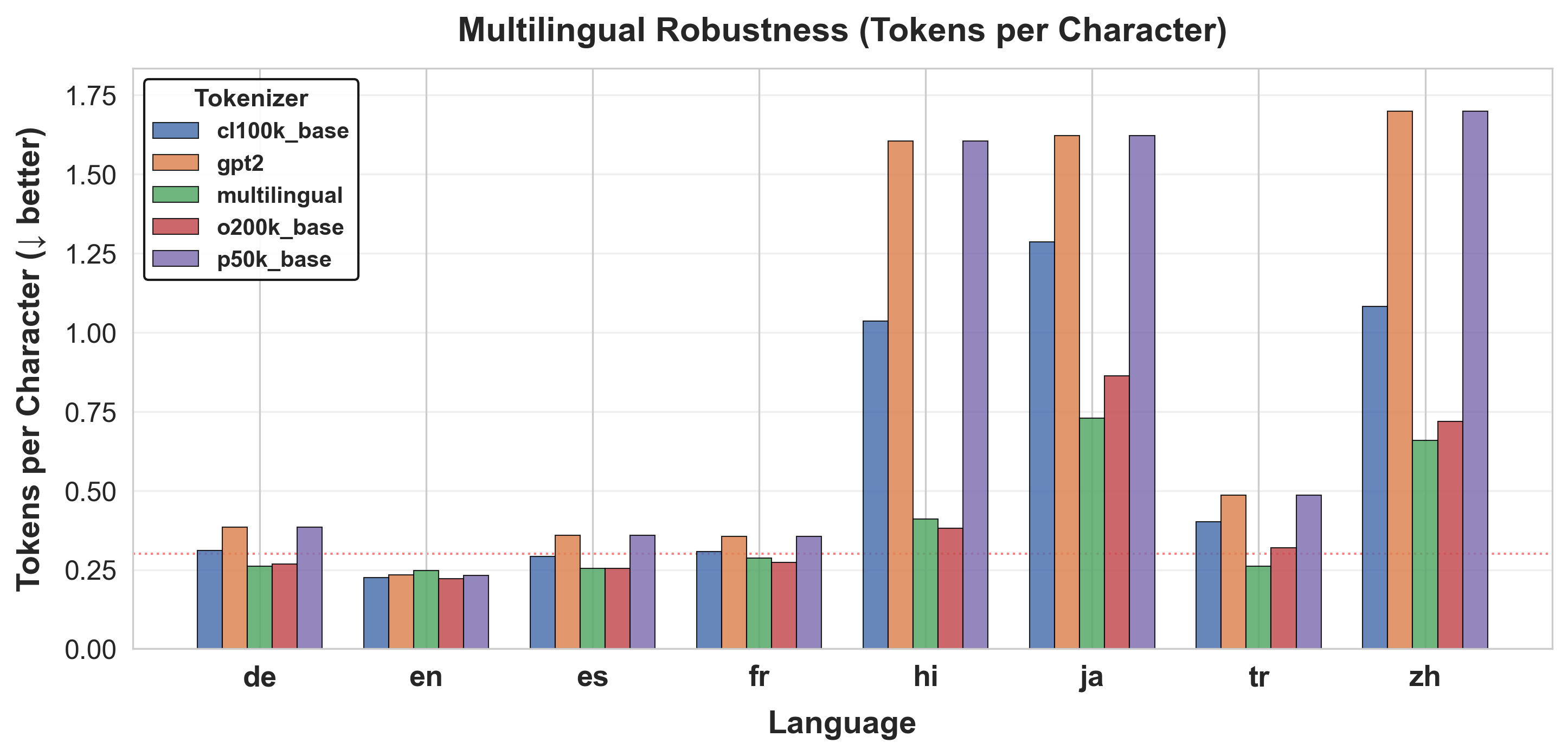}
    } \hspace{-0.45cm}

    \caption{Analyzing the GPT-family Tokenizers}
    \label{fig:scatter}
    \vspace{-0.4cm}
\end{figure}

The x-axis (zstd bits/char) reflects the intrinsic compressibility of each domain: with code being highly structured ($\sim$3 bpc), news moderate ($\sim$4.6 bpc), \& math the most complex ($\sim$5.5 bpc). The y-axis (tokens/char) calibrates how well each baseline tokenizes. On news/math, all baselines yield very similar performance, with \texttt{o200k\_base} being slightly more compact. In contrast, on code, the two newer tokenizers sit at the bottom implying better compression. This also suggests that tokenizer design (and a larger training vocabulary size) matters most in highly structured domains like code, and much less in standard English or math-heavy text.

We next probe \textbf{cross-lingual robustness} of these tokenizers. We consider eight languages: German (de) , English (en), Spanish (esp), French (fr), Turkish (tr), Hindi (hi), Japanese (ja), and Chinese (zh). For each language we draw up to 2 million characters of web text from two open-source corpora: C4~\cite{raffel2020exploring} and Oscar~\cite{ortiz-suarez-oscar} datasets, and evaluate the GPT tokenizers alongside a multilingual tokenizer (XLM-RoBERTa)~\cite{xlm-roberta} trained over 100+ languages. For every language–tokenizer pair, we compute the average tokens per character (Fig.~\ref{fig:scatter}b).

Across the Latin derivatives (de/en/es/fr), all baselines achieve performance close to the multilingual baseline, with newer tokenizers being slightly more compact. For Turkish, the baselines start to fragment words more than the multilingual tokenizer, but the gap remains moderate. In contrast, for Hindi, Japanese, and Chinese, most baselines produce several times more tokens per character, indicating severe over-segmentation of non-Latin scripts. The multilingual tokenizer and \texttt{o200k\_base} sit much closer in the plot, significantly narrowing this gap. Overall, this experiment highlights that cross-lingual robustness is highly tokenizer-dependent, and that the newer \texttt{o200k\_base} design moves LLM tokenization closer to a multilingual, script-aware regime.

\section{Compressibility of Learned Tokenizers}
\label{sec:learned}
We conduct a systematic comparison of four tokenizer families across multiple domains and training sizes, to understand how tokenization affects both compression efficiency and local predictability of the resulting token sequences.

We compare four tokenization schemes that cover the standard subword paradigms: BPE~\cite{gage1994new}, Unigram~\cite{kudo2018subword, kudo2018sentencepiece}, WordPiece~\cite{song2021fast}, WordLevel~\cite{chen1996}. All tokenizers are \textbf{trained from scratch} for our experiments. Brief descriptions and practical considerations for each tokenizer are provided in Appendix~\ref{app:tokenizers}. The evaluation spans four domains drawn from publicly available HuggingFace datasets. For natural language, we use web text from the C4 dataset~\cite{raffel2020exploring}, considering the English, Turkish, and Chinese splits. These languages represent complementary linguistic regimes: English is morphologically simpler and whitespace-segmented, Turkish is agglutinative and morphologically rich, and Chinese uses a non-Latin script without explicit word boundaries. Additionally, we evaluate on multilingual source code sampled from Bigcode's Starcoder dataset~\cite{li2023starcoder}, which constitutes a structurally distinct domain characterized by high symbol diversity and different compositional patterns. \looseness=-2

For each domain, we stream characters from the dataset until a fixed maximum length is reached. The resulting text stream is then partitioned into two components: a training portion, consisting of an initial prefix truncated to one of several target sizes, and a test portion, defined as the final 10 million characters of the stream held out exclusively for evaluation. This procedure ensures that, within each domain, all tokenizers are evaluated on an identical test slice, independent of the size or composition of their respective training corpora.

\textit{Tokenizers Training \& Evaluation:} We fix the tokenizer vocabulary sizes to $|\mathcal{V}|= \{16000, 64000\}$ for all tokenizer families and domains. For each domain, we train tokenizers on progressively larger prefixes of the training text, with total character counts ranging from approximately $10^3$ to $10^8$. This allows us to study how tokenizer performance scales with training data. All tokenizers are trained using a common \textbf{pipeline}: i) for \textbf{preprocessing}, we apply NFKC (standard Unicode norm.), a whitespace-based pre-tokenization, and use a consistent set of special tokens across tokenizers (e.g., \texttt{<pad>}, \texttt{<unk>}), and ii) for \textbf{training} across each (domain, tokenizer, training size) config., we train a new tokenizer on the corresponding text. %

\textit{Evaluation Metrics:} We evaluate each trained tokenizer on the held-out 10M-character test slice using two classes of metrics: a) compression ratio and b) empirical $k$-gram entropies (unigram through 5-gram). All metrics are computed on the \textbf{token sequence} produced from the test text.

\subsection*{A. Compression Ratio:}

\begin{figure}[t!]
    \centering
    \vspace{-0.4cm}
    \subfigure[English]{
        \includegraphics[width=0.48\linewidth]{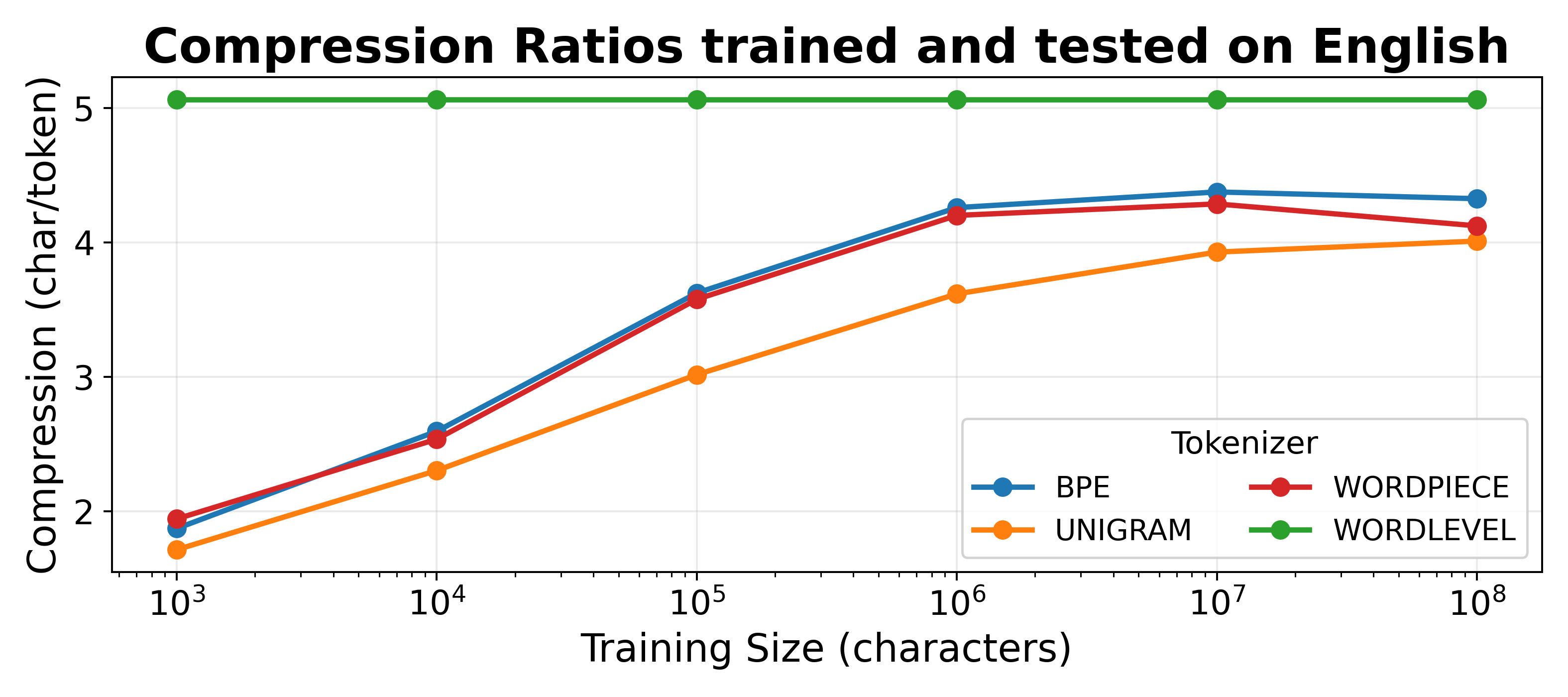}
        \vspace{-1cm}
    } \hspace{-0.3cm}
    \subfigure[Code]{
        \includegraphics[width=0.48\linewidth]{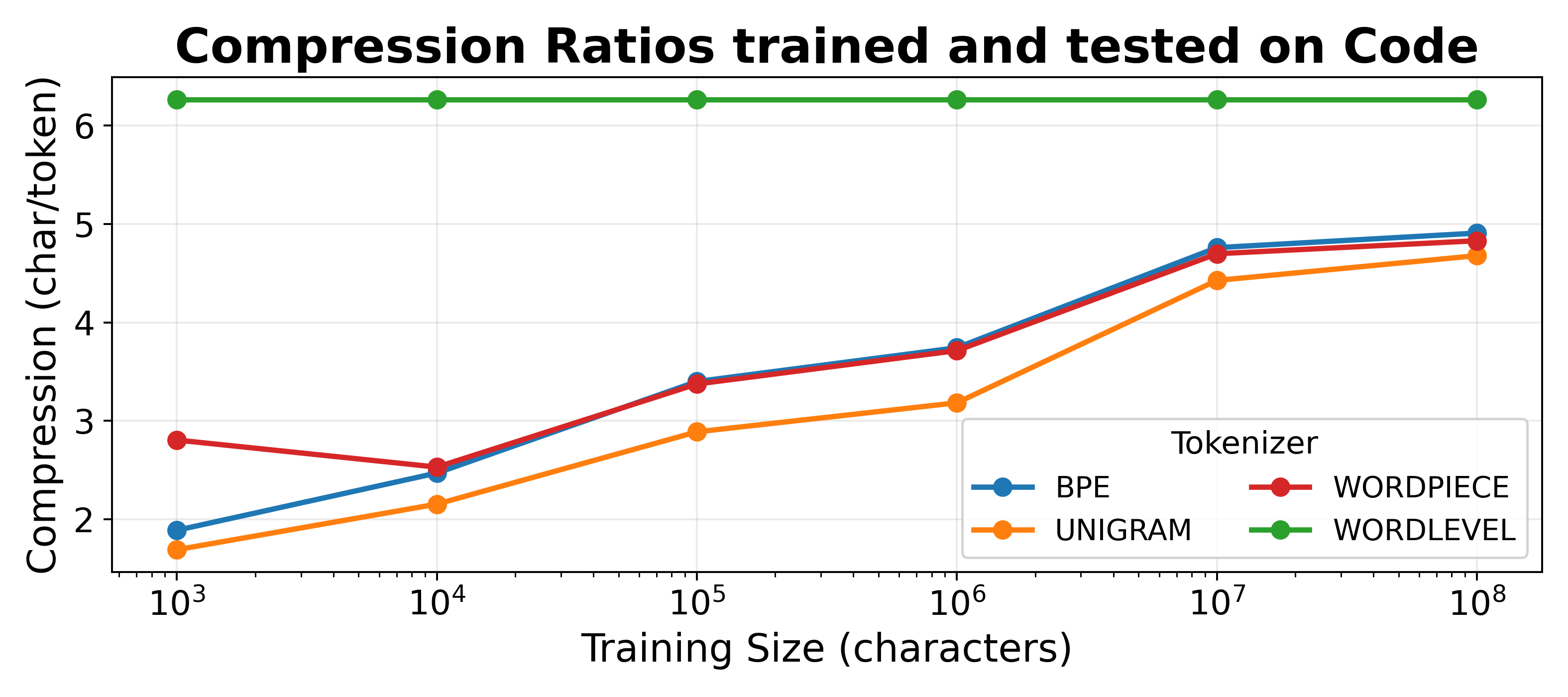}
        \vspace{-1cm}
    } \hspace{-0.3cm}
    
    \vspace{-0.2cm}
    
    \caption{Compression ratios across domains (vocab. size = 16k)}
    \vspace{-0.4cm}
    \label{fig:compression-grid16k}
\end{figure}

\begin{figure}[t!]
    \centering

    \subfigure[English]{
        \includegraphics[width=0.48\linewidth]{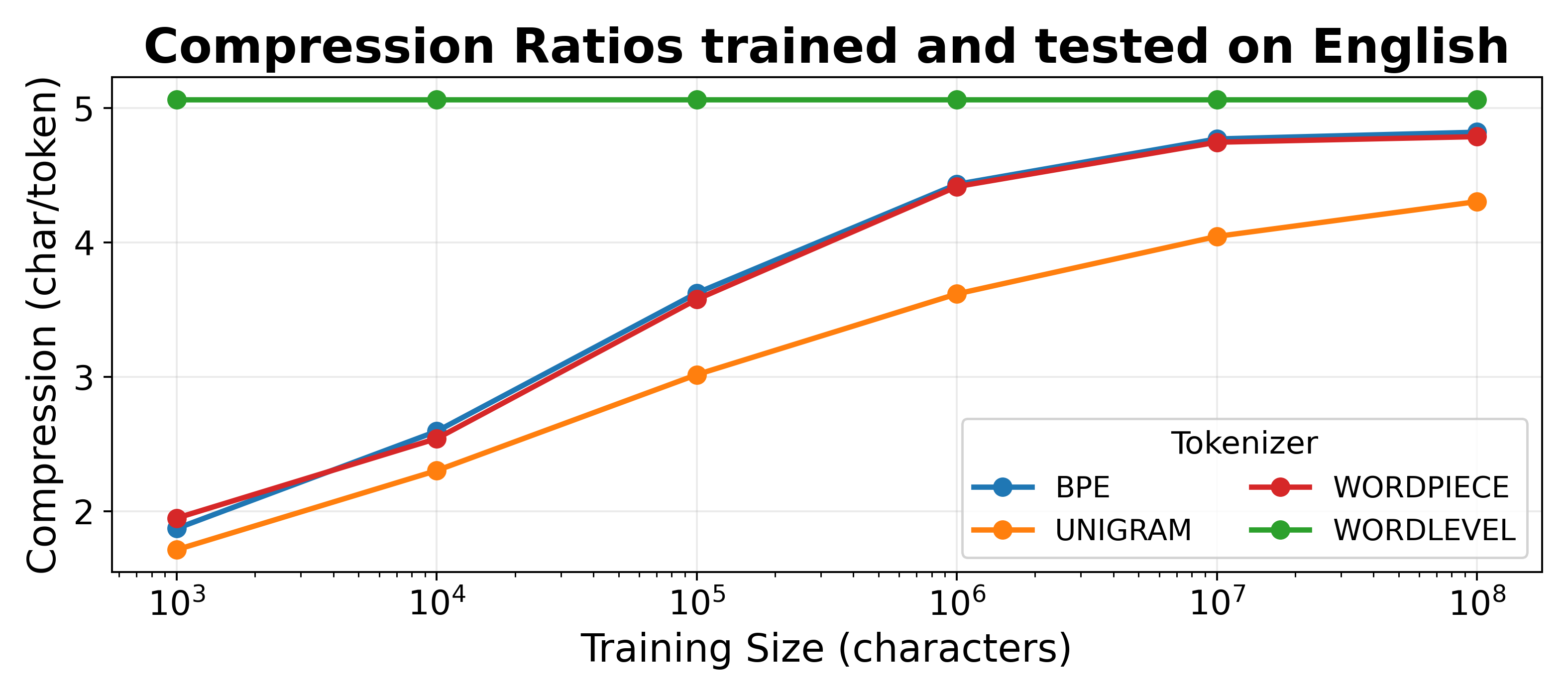}
    } \hspace{-0.3cm}
    \subfigure[Code]{
        \includegraphics[width=0.48\linewidth]{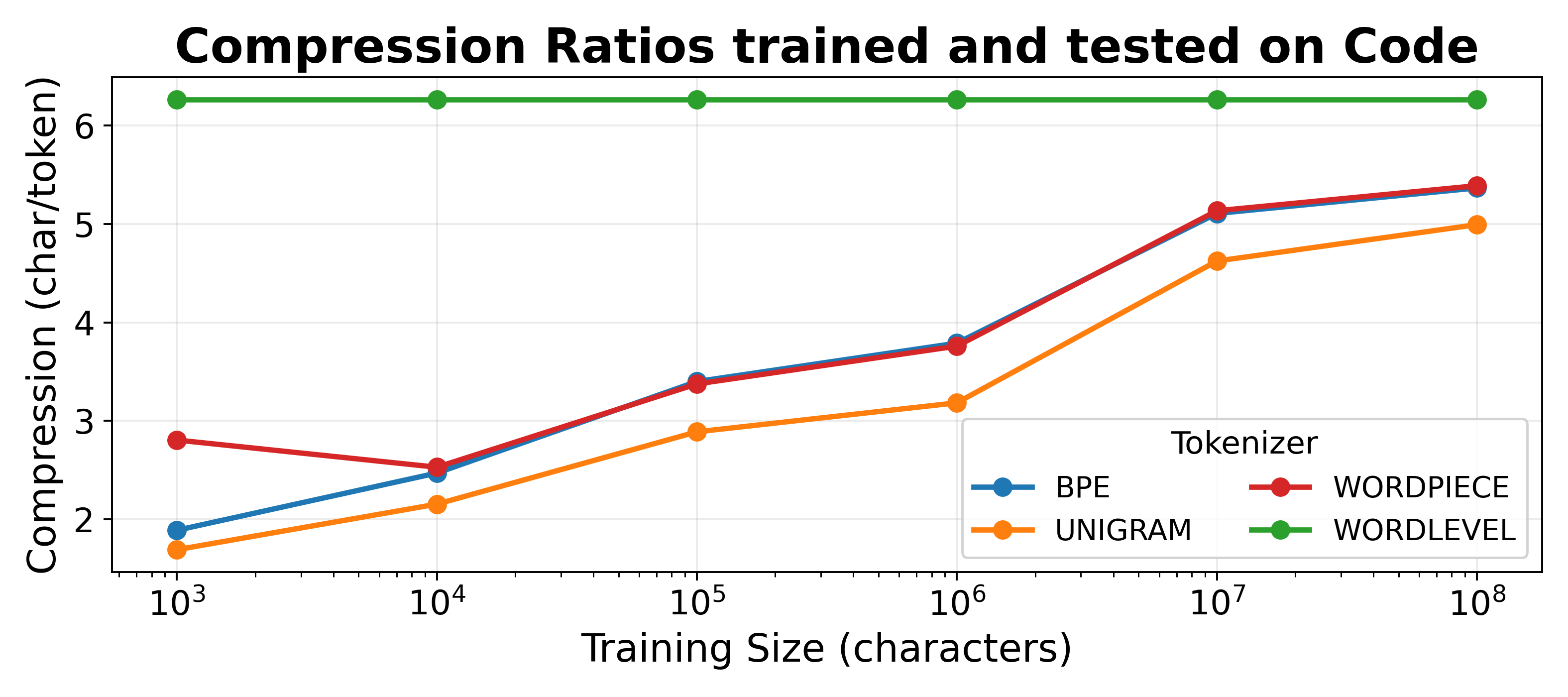}
    } \hspace{-0.3cm}
    \vspace{-0.1cm}

    \caption{Compression ratios across domains (vocab. size = 64k)}
    \vspace{-0.5cm}
    \label{fig:compression-grid64k}
\end{figure}

For a tokenizer $T$ and a test corpus $\mathcal{D}$, let $|x|_{\text{UTF-8}}$ denote the length (in bytes) of a string $x$ under UTF-8, and let $|T(x)|$ denote the number of tokens produced by $T$ on $x$. We define the \emph{compression-ratio} of $T$ on $\mathcal{D}$ as:
\begin{equation*}
\mathrm{CR}(T; \mathcal{D}) = \frac{\sum_{x \in \mathcal{D}} |x|_{\text{UTF-8}}}{\sum_{x \in \mathcal{D}} |T(x)|},
\end{equation*}
i.e., the average number of UTF-8 bytes per token on the test corpus. \textit{Larger} values of $\mathrm{CR}$ correspond to \textit{more compressive} tokenizations (fewer tokens per character) under this metric.

\noindent
~\autoref{fig:compression-grid16k},~\ref{fig:compression-grid64k} show the compression performance of these trained tokenizers for varying training size with vocabulary sizes 16k and 64k. Trivially, the compression ratio achieved by WordLevel tokenizer remains constant with training size. Interestingly, we see that in the case of English (vocabular size = 16k), the performance of BPE and WordPiece takes a hit once the training size increases from $10^7$ to $10^8$ characters.

\begin{figure}[t!]
    \centering
    \vspace{-0.215cm}
    \subfigure[$k$-gram Entropy ($H_k$)]{
        \includegraphics[width=1\linewidth,
          trim=0 0.6cm 0 1.2cm, clip]{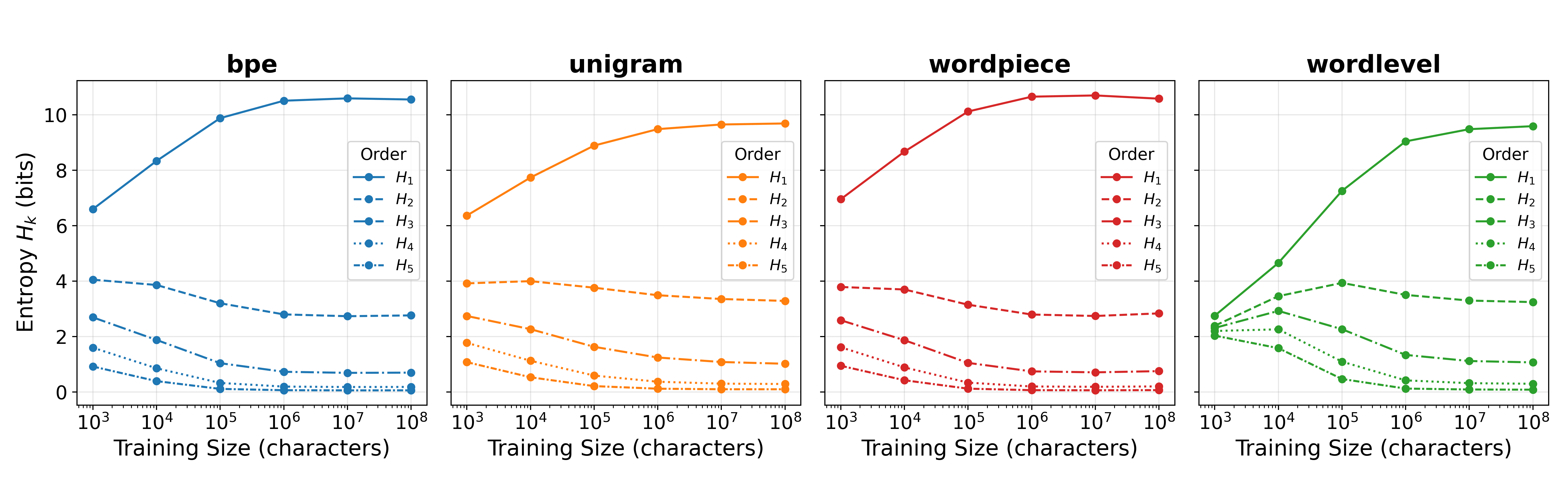}
    } 
    \vspace{-0.1cm}
    \subfigure[$k$-gram Entropy Rate ($H_k$ $\times$ token/char)]{
        \includegraphics[width=1\linewidth,
          trim=0 0.6cm 0 1.2cm, clip]{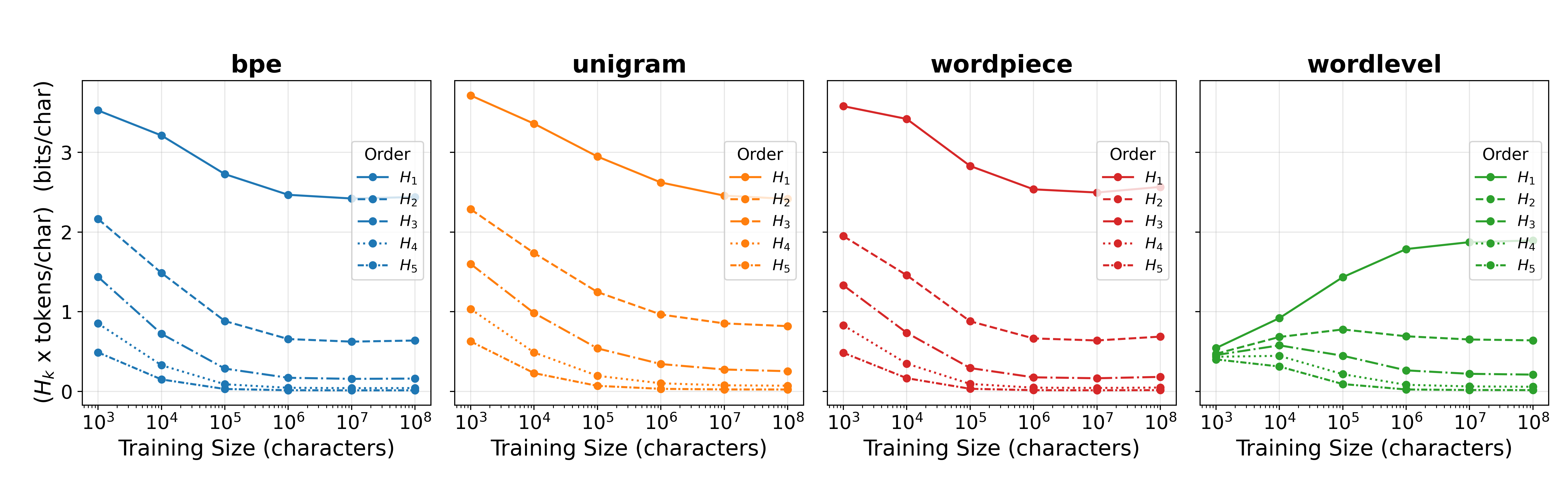}
    } 
    \vspace{-0.1cm}
    \caption{Tokenizer $k$-gram entropy results trained \& tested on English for vocabulary size 16k. \looseness=-2}
    \vspace{-0.5cm}
    \label{fig:kthorderEnglish}
\end{figure}

However, for vocabulary size of 64k, the compression ratio increases monotonically. We interpret this \textbf{not} as classical sample overfitting, but as a \textit{capacity-limited} universal coding effect: with a small vocabulary the learned dictionary that is optimal for the large training corpus need not minimize cross-entropy on our fixed test source, whereas with a larger vocabulary this capacity constraint is relaxed~\cite{feder2025information} (see Appendix~\ref{app:modelcapacity} for further discussion). Furthermore, on Chinese (in~\autoref{fig:compression-grid16k-app}, ~\ref{fig:compression-grid64k-app}), we observe degradation with increasing training size in all cases except BPE, consistent with its large unique-character set stressing vocabulary allocation in non-BPE tokenizers.

To better isolate this behavior, we repeated the experiments using a substantially larger vocabulary of 500k tokens for both Chinese and Chinese-Latin (Pinyin-style transliteration) (Figure \ref{fig:compression-grid64k-app-2}). At this vocabulary scale, BPE shows the expected monotonic improvement in compression, whereas WordLevel and Unigram exhibit a characteristic dip: performance initially worsens before improving again as the vocabulary becomes large enough. For code, however, the performance of these tokenizers increases monotonically with the training size.

\subsection*{B. $k$-gram Entropies:} 
Let $T = (t_1,\ldots,t_n)$ be the tokenized test sequence and $\hat{p}(t)$ be the empirical frequency of token $t$. The unigram entropy can be computed using: $H_1 = - \sum_{t} \hat{p}(t)\,\log \hat{p}(t)$.

To assess local predictability and the strength of short-range dependencies induced by the tokenizer, we compute empirical conditional entropies of order $k = 2,\ldots,5$. For each order $k$:
\begin{itemize}
\item We construct counts of all observed length-$k$ token tuples
$(t_{i-k+1},\ldots,t_i)$ in the test sequence.
\item From these counts, we derive empirical conditional distributions $\hat{p}(t_i | t_{i-k+1}^{i-1})$.
\end{itemize}

We then compute the empirical $k$-gram entropy as:
\begin{equation*}
\widehat{H}_k = \frac{1}{n}\sum_{i=1}^{n} \bigl[-\widehat{p}\bigl(t_i| t_{i-k+1}^{i-1}\bigr) \log \widehat{p}\bigl(t_i| t_{i-k+1}^{i-1}\bigr)\bigr].
\end{equation*}
\noindent
Figure~\ref{fig:kthorderEnglish}a reports token-level conditional entropies on English as a function of tokenizer family and training size. A consistent pattern emerges: as training increases from \(10^3\) to \(10^8\) characters, the unigram entropy \(H_1\) grows, reflecting a richer and more uniform token distribution, while higher-order entropies \(H_k\) for \(k \geq 2\) steadily decrease. For example, with BPE, \(H_1\) rises from roughly \(7\) to \(10.5\) bits, whereas \(H_4\) and \(H_5\) fall from about \(1\) bit to nearly zero. Unigram, WordPiece, and WordLevel show similar trends, with slightly larger residual \(H_k\) for WordLevel. We also see a similar trend in other languages and domains as shown in Figures~\ref{fig:ent1}-\ref{fig:ent5z}.

Figure~\ref{fig:kthorderEnglish}b complements this view by reporting the entropy rate (bits/character). Although the unigram \(H_1\) increases with training, tokens/char decreases, so the net rate drops. Moreover, higher-order rates decrease even more sharply, indicating a more predictable token stream in context. These trends are consistent with the gains from applying universal compressors after tokenization observed in Section~\ref{sec:lzontokenization}.
\looseness=-2

Overall, conditioning on some token context leaves less than 1 bit of uncertainty, indicating that most local structure in text is absorbed by the tokenization. This aligns with the theory of~\cite{towardTokenizationTheory2024}, showing that appropriate tokenization can capture low-order dependencies, so that the downstream transformer can devote more of its capacity to modeling longer-range structure. Additional results for other domains and vocabulary sizes appear in Figures~\ref{fig:ent1}–\ref{fig:gpt-tok} (including $k$-gram entropy analysis for pre-trained GPT tokenizers in Fig.~\ref{fig:gpt-tok}). \looseness=-2

\subsection*{C. Tokenizers Under Train–Test Domain Mismatch}

\begin{figure}[t!]
    \centering
    \vspace{-0.35cm}
    \subfigure[Turkish]{
        \includegraphics[width=0.48\linewidth]{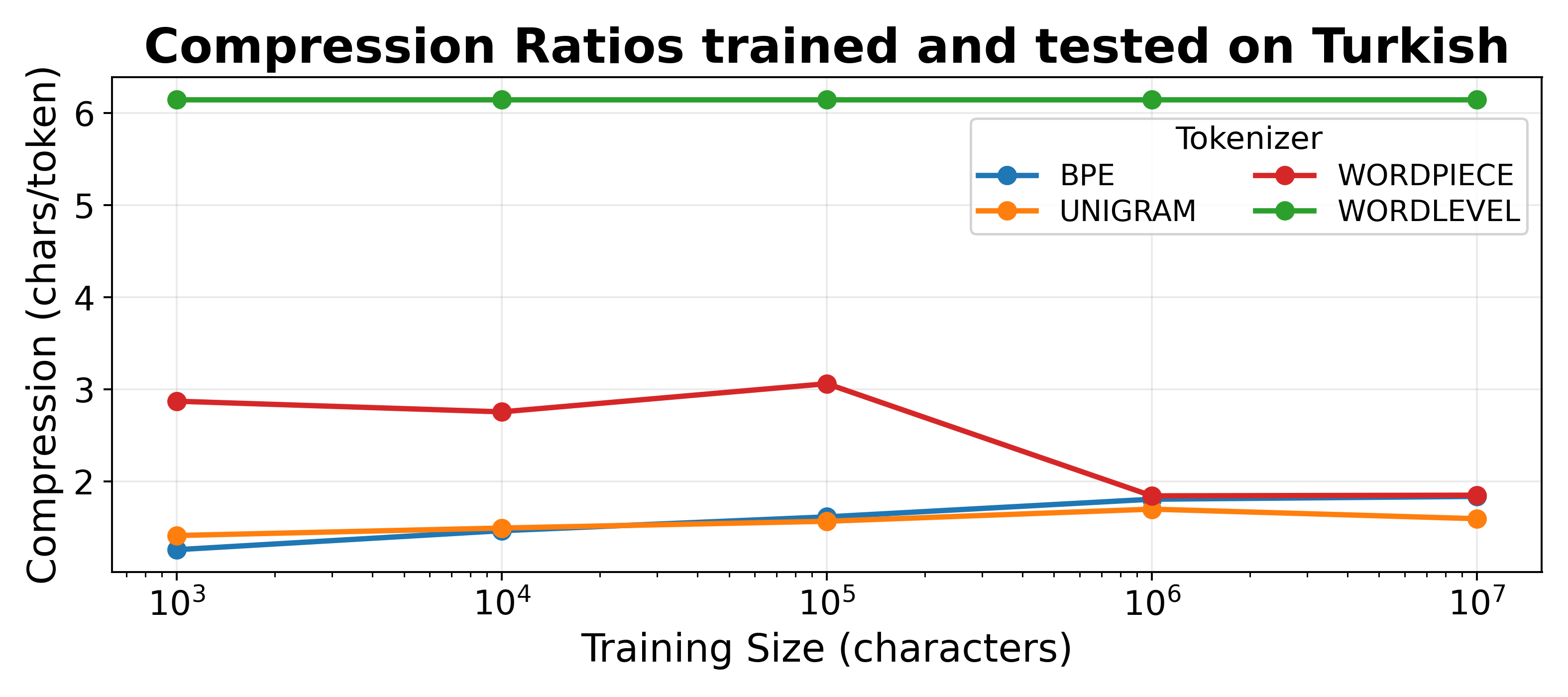}
        \vspace{-1cm}
    } \hspace{-0.3cm}
    \subfigure[Code]{
        \includegraphics[width=0.48\linewidth]{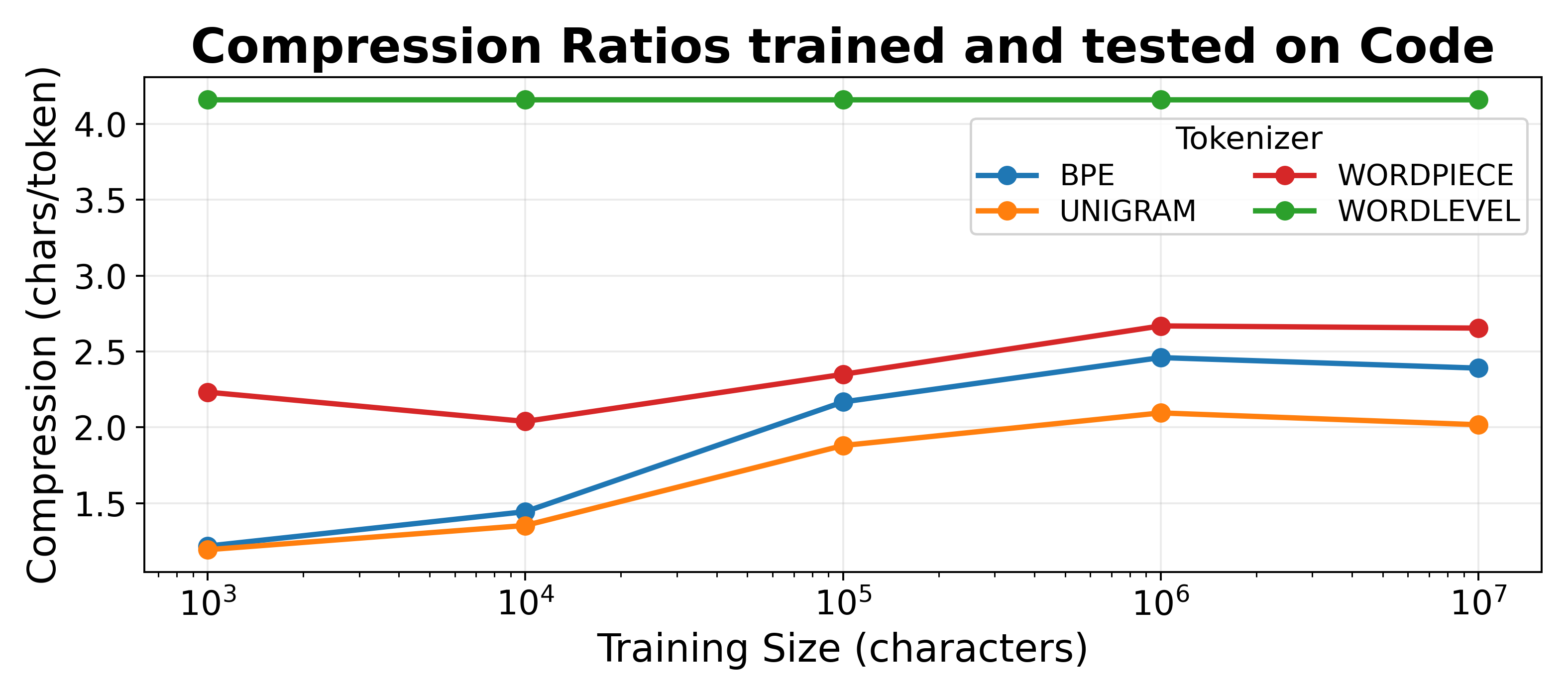}
        \vspace{-1cm}
    } \hspace{-0.3cm}

    \vspace{-0.2cm}
    \caption{Compression ratios in domain mismatch (vocab. size = 16k).
    Trained on English, tested on (a) Turkish, (b) Code. \looseness=-2}
    \label{fig:compression-domainmismatch}
    \vspace{-0.5cm}
\end{figure}

\begin{figure}[t!]
\centering
    \includegraphics[width=1\linewidth]{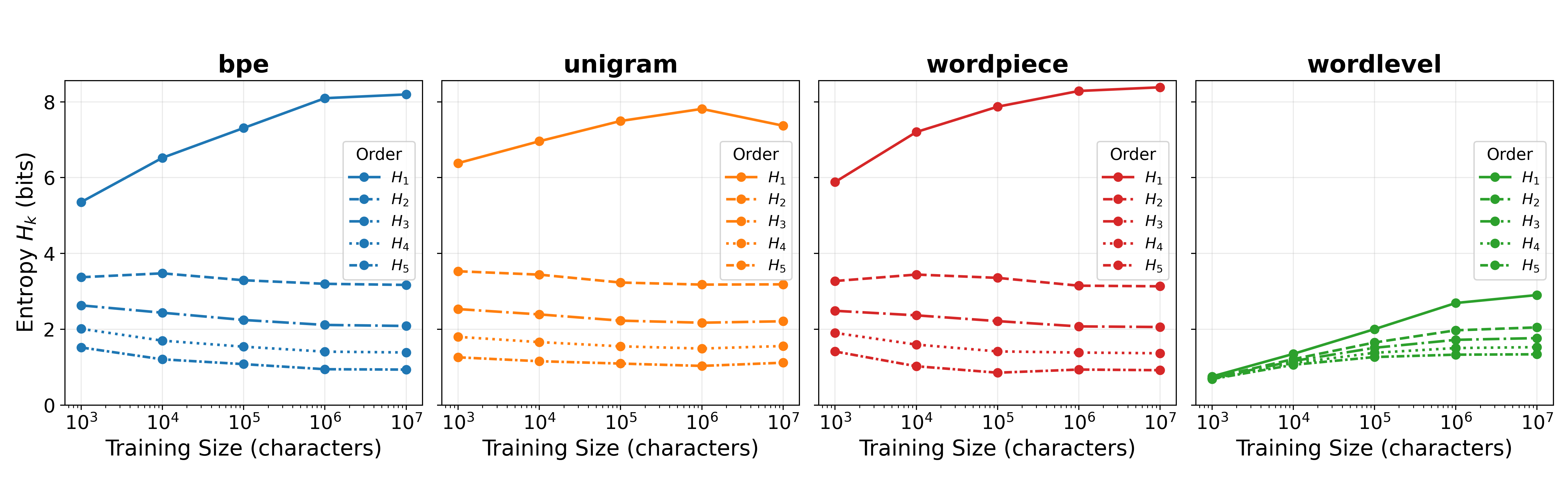}
    \vspace{-0.7cm}
    \caption{$k$-gram entropies in domain mismatch (vocab. 16k)}
\label{fig:distribution_code}
\vspace{-0.5cm}
\end{figure}

We next evaluate tokenizers under training–test domain mismatch. When a tokenizer is trained on English but tested on Turkish, Code, or Chinese, its compression performance does not consistently improve with larger English training sets (Fig.~\ref{fig:compression-domainmismatch}). Moreover, the empirical conditional entropies for $k>1$ remain far from zero (there is always some offset), indicating that these tokenizers rely heavily on properties of their training corpus and can degrade substantially when applied to unaligned domains. These phenomena are briefly illustrated below in Figures~\ref{fig:compression-domainmismatch} and \ref{fig:distribution_code}.

\section{Tokenization and Universal Compressors}
In this section, we investigate how learned tokenization interacts with universal compressors using a two-stage pipeline below, viewing tokenization as a transform:
\[
\text{text}
  \xrightarrow{\text{tokenizer }\tau_K}
\text{token sequence}
  \xrightarrow{\text{LZ / CTW / etc.}}
\text{bitstream}.
\]

\subsection{LZ Compression on Tokenized Sequences}
\label{sec:lzontokenization}
We study whether learned tokenization can make text more compressible for off-the-shelf LZ-style compressors. For a given corpus we compare \emph{raw-LZ}, where we apply gzip/lzma/zstd~\cite{lz77, collet2018zstandard, lzma} directly to UTF-8 bytes, to a two-stage pipeline where we first tokenize, run LZ on a derived representation (where the tokenizer is viewed as a transform):
\vspace{-0.4cm}
\begin{align*}
    \text{Raw-LZ:} \text{ } & \text{UTF-8 bytes} \xrightarrow{\text{gzip / lzma / zstd}} \text{bits}, \\
    \text{Token-LZ:} \text{ } & \text{UTF-8} \xrightarrow{\text{tokenizer}} \text{token IDs} \xrightarrow{\text{16-bit ints}} \xrightarrow{\text{gzip/lzma/zstd}} \text{bits},
\end{align*}

Table~\ref{tab:lz-token-ids} reports bits per character (bpc) when we compress
the 16-bit token ID sequence. On raw UTF-8, LZMA achieves $\approx 2.55$\,bpc
(on this blocklength), with gzip and zstd at $\approx 3.07$ and
$2.60$\,bpc respectively. After tokenization, all three tokenizers improve compression by roughly $10$–$20\%$, with the Unigram model consistently giving the
most compressible sequences, BPE slightly worse, and WordPiece worst. This illustrates that a simple ``tokenizer $\rightarrow$ LZ'' pipeline outperforms raw LZ, suggesting that learned tokenization acts as a beneficial finite-sample transform that simplifies structure prior to (universal) compression. \looseness=-2

\begin{table}[ht]
\centering
\small
\setlength{\tabcolsep}{6pt}
\resizebox{\columnwidth}{!}{%
\begin{tabular}{lcccc}
\hline
\textbf{Compressor} & \textbf{Raw bytes} & \textbf{BPE IDs} & \textbf{Unigram IDs} & \textbf{WordPiece IDs} \\
\hline
\textbf{gzip} & 3.07 & 2.44 (-20\%) & 2.42 (-21\%) & 2.50 (-19\%) \\
\textbf{lzma} & 2.55 & 2.13 (-16\%) & 2.11 (-17\%) & 2.17 (-15\%) \\
\textbf{zstd} & 2.60 & 2.31 (-11\%) & 2.27 (-13\%) & 2.36 (-9\%)  \\
\hline
\end{tabular}%
}
\vspace{2pt}
\caption{bits per character (bpc) on C4 dataset when applying LZ directly to raw UTF-8 text vs.\ tokenized text.}
\label{tab:lz-token-ids}
\vspace{-0.4cm}
\end{table}

\subsection{LZ-Aware BPE}
Motivated by the observation that reduction in token-level entropy does not necessarily improve downstream compression, we consider a proof-of-concept \emph{LZ-aware BPE} procedure that explicitly optimizes merges for a universal-compressor objective. This conceptually aligns with recent work on tokenizer design tailored to a fixed downstream model~\cite{towardTokenizationTheory2024}.
\noindent
\paragraph*{Algorithm sketch}
Standard BPE initializes with a byte vocabulary and repeatedly merges the most frequent adjacent token pair until reaching a target vocabulary size. In contrast, LZ-aware BPE selects each merge by greedily minimizing the gzip-compressed length of a held-out validation stream. Concretely, we start from the \emph{byte alphabet} (initial vocabulary size \(|V|=256\)), represent both training and validation text as sequences of byte-token IDs, and then iterate:
\begin{enumerate}
    \item Compute frequencies of adjacent token pairs on the \emph{training} token stream.
    \item Form a candidate set consisting of the top-\(K\) most frequent pairs (we use \(K=50\)).
    \item For each candidate merge, simulate applying the merge to the \emph{validation} stream, pack the resulting token IDs as a 16-bit byte stream, compress with gzip, and record the compressed size (bytes).
    \item Commit the merge achieving the smallest validation compressed size, update the tokenization (and hence the training/validation token streams), and repeat until the target \(|V|\) is reached.
\end{enumerate}

\begin{figure}[t]
    \centering
    \includegraphics[width=0.97\linewidth]{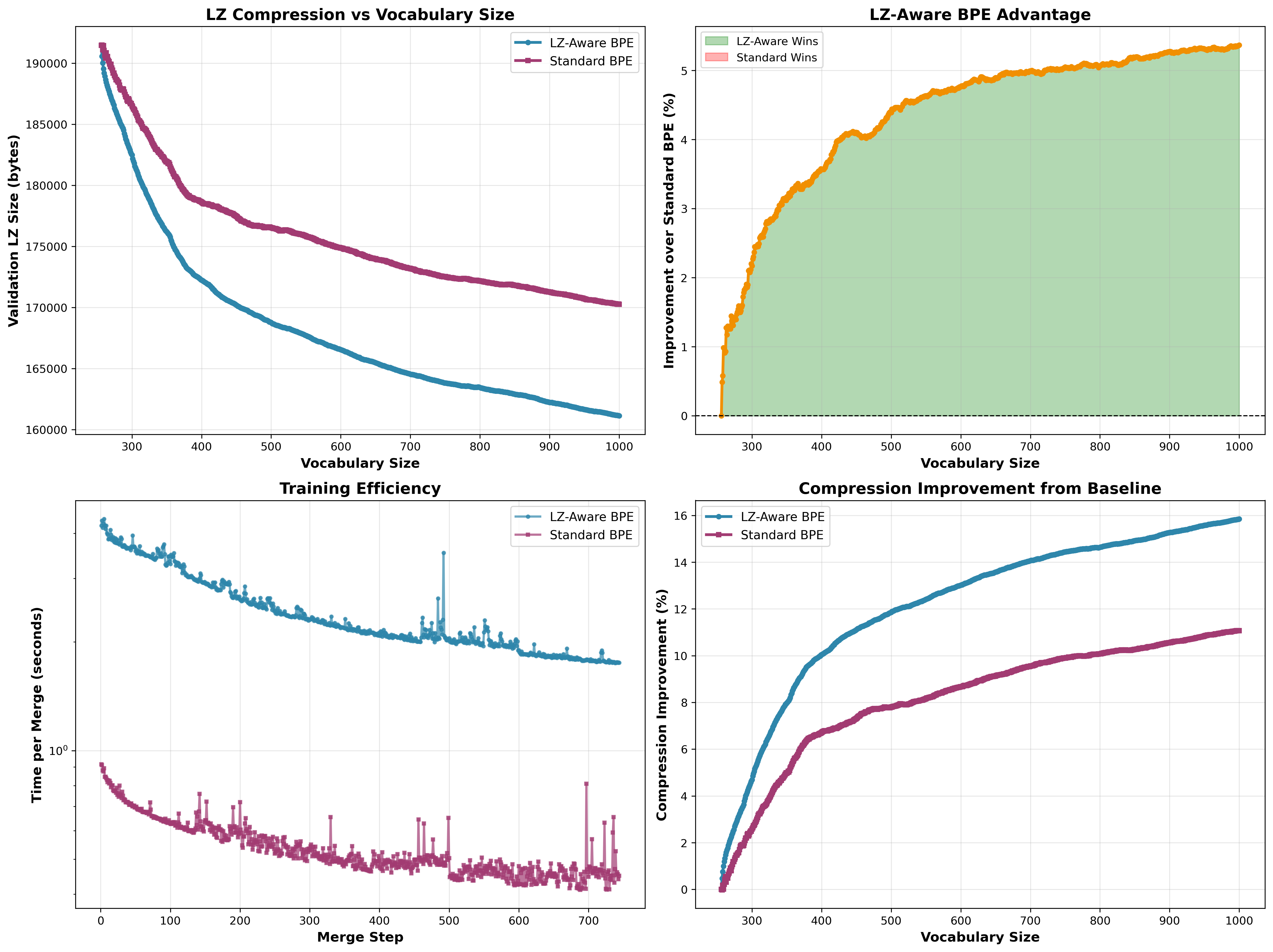}
    \caption{\small LZ-Aware BPE results}
    \label{fig:lzawarebpe}
    \vspace{-0.5cm}
\end{figure}

\noindent
\paragraph*{Experimental setup and results}
We train on English web text from C4 dataset \cite{raffel2020exploring}, using a total budget of \(5\times 10^6\) characters and reserving \(20\%\) for validation. Due to the cost of evaluating \(K\) candidate merges via recompression at each step, we report results for vocabularies in the range \(|V|\in[256,1024]\). We also train a standard frequency-based BPE tokenizer on the same data and sweep the same vocabulary sizes. We evaluate using gzip on (i) raw UTF-8 bytes and (ii) packed token-ID streams, reporting compressed size and derived rates. Across the entire sweep, LZ-aware BPE yields smaller validation gzip size than standard BPE (top-left, Fig.~\ref{fig:lzawarebpe}). Relative to the byte-level baseline (\(|V|=256\)), LZ-aware BPE achieves an overall compression improvement of \(\approx 15.8\%\) by \(|V|\approx 1024\), compared to \(\approx 11.1\%\) for standard BPE (bottom-right, Fig.~\ref{fig:lzawarebpe}). This gain comes at a computational cost: the per-merge time increases by several-fold (median \(\sim\!2\)s per merge for LZ-aware vs.\ \(\sim\!0.4\)s for standard; bottom-left). We leave a study of its impact on downstream language-modeling (e.g., perplexity) to future work.

\section{Tokenization Through a Channel Lens}
\label{sec:tok-chan}
Beyond viewing tokenizers as compressors, we also adopt the \emph{noiseless channel} perspective of~\cite{zouhar2023tokenization}. Formally, a tokenizer $T$ induces a deterministic mapping $T: \mathcal{X}^* \to \mathcal{V}^*$, where $\mathcal{X}$ is the input alphabet (UTF-8 bytes or characters) and $\mathcal{V}$ is a finite token vocabulary of size $|\mathcal{V}| = K$. Interpreting tokens as symbols transmitted over a noiseless $K$-ary channel, the per-token capacity is
$C_{\text{token}}=\log_2 K$ bits/token, while the empirical information carried on corpus $D$ is the unigram entropy
$H_1(T;D)$. We therefore define the \emph{capacity utilization}
\begin{equation}
\eta(T;D)\triangleq \frac{H_1(T;D)}{\log_2 K}.
\end{equation}
Following~\cite{zouhar2023tokenization}, we also consider a Rényi analogue: for $\alpha>1$,
\[
\eta_\alpha(T;D) \;\triangleq\; \frac{H_\alpha(T;D)}{\log_2 K},
\]
which places greater weight on the head of the token distribution (e.g., $\alpha { = } 2$ corresponds to collision entropy). This channel view helps tie together several empirical observations.

\begin{figure}
    \centering
    \includegraphics[width=0.75\linewidth]{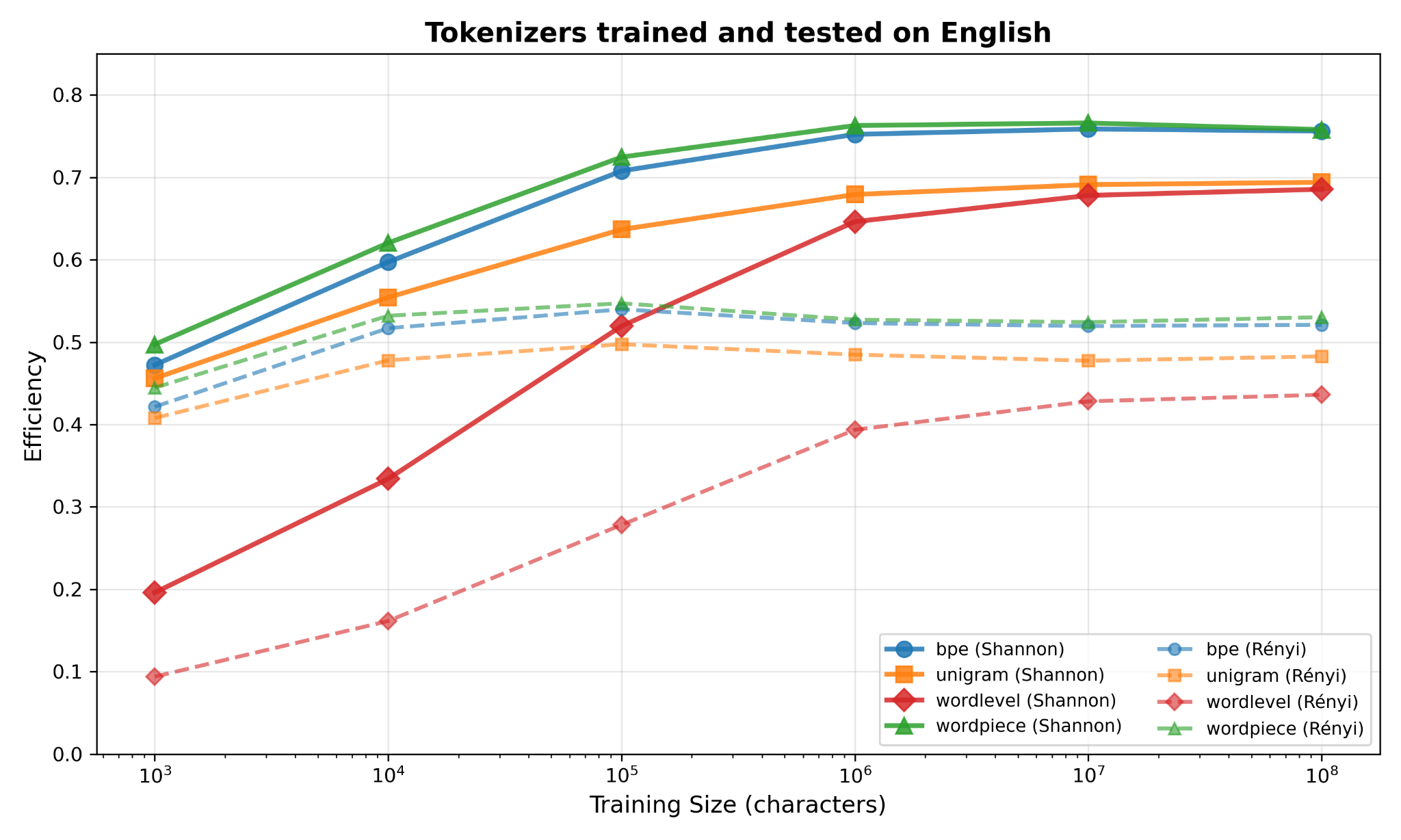}
    \caption{Capacity Utilization vs. Training Size}
    \label{fig:cap}
    \vspace{-0.5cm}
\end{figure}

On English with $K = 16k$, $\eta$ increases rapidly with tokenizer training size and then saturates (Fig.~\ref{fig:cap}), indicating that modest training data already captures most attainable \emph{marginal} vocabulary usage. Across families, subword tokenizers achieve high utilization while simultaneously reducing higher-order conditional entropies $H_k$, consistent with tokenization shifting short-range regularities into the representation. Notably, WordLevel can attain comparable Shannon utilization at large training sizes, yet leaves larger $H_k$ values, indicating more uniform vocabulary usage without absorbing low-order structure. In contrast, $\eta_2$ can plateau or decline even when $\eta$ increases, showing a growing probability mass concentration among a small set of very frequent tokens, an effect that is not visible from sequence length alone. Empirically, additional training data can raise $H_1$ by populating the long tail while simultaneously lowering $H_2$ via increased head concentration among very frequent tokens. These utilization measures provide a compact complement to our compression and $k$-gram analyses, and motivate future work probing how $\eta$/$\eta_\alpha$ correlate with intrinsic modeling difficulty (e.g., perplexity or bits/byte under small LMs). Further details can be found in Appendix~\ref{app:tok-chan}. \looseness=-2

\section{Conclusion}
\label{sec:concld}

Our experiments reveal that tokenizers function as structured compressors whose compression performance and induced token-stream statistics are determined by domain, vocabulary size and training scale. Pretrained GPT-family tokenizers exhibit relatively stable compression behavior on English/Math/Code, but are brittle in multilingual settings, often over-segmenting non-Latin scripts. Learned tokenizers expose trade-offs across training scale and vocabulary size: they absorb short-range structure, yet the benefits are distribution-dependent and can degrade under train--test mismatch. These findings highlight the importance of script-aware preprocessing and domain-aligned training/evaluation in tokenizer design. Overall, our results support an information-theoretic view of tokenization as a finite-alphabet channel that mediates trade-offs between token count, induced structure, and robustness. A key direction for future work is to develop a more explicit channel-based perspective for tokenizer design and evaluate its effectiveness in downstream language modeling performance (e.g., perplexity). \looseness=-2

\newpage

{\footnotesize
\bibliographystyle{ieeetr}
\bibliography{references}

@article{goose2024compressing,
  title   = {Training LLMs over Neurally Compressed Text},
  author  = {{Lester et al.}},
  journal = {arXiv preprint arXiv:2404.03626},
  year    = {2024},
  url     = {https://arxiv.org/abs/2404.03626}
}

@article{tokenizercontrol2024,
  title   = {Tokenization Is More Than Compression},
  author  = {{Schmidt et al.}},
  journal = {arXiv preprint arXiv:2402.18376},
  year    = {2024},
  url     = {https://arxiv.org/abs/2402.18376}
}

@manual{lzma,
  author       = {Igor Pavlov},
  title        = {{LZMA — Lempel–Ziv–Markov chain Algorithm}},
  organization = {7-Zip.org},
  year         = {1998},
  note         = {Original 7-Zip implementation and SDK; LZMA algorithm documentation},
  url          = {https://www.7-zip.org/},
}

@inproceedings{young2025radio,
title={Radio: Rate{\textendash}Distortion Optimization for Large Language Model Compression},
author={Sean I. Young},
booktitle={Forty-second International Conference on Machine Learning},
year={2025},
url={https://openreview.net/forum?id=ifnxXCCEiM}
}

@inproceedings{zouhar2023tokenization,
    title = "Tokenization and the Noiseless Channel",
    author = "Zouhar, Vil{\'e}m  and
      Meister, Clara  and
      Gastaldi, Juan  and
      Du, Li  and
      Sachan, Mrinmaya  and
      Cotterell, Ryan",
    editor = "Rogers, Anna  and
      Boyd-Graber, Jordan  and
      Okazaki, Naoaki",
    booktitle = "Proceedings of the 61st Annual Meeting of the Association for Computational Linguistics (Volume 1: Long Papers)",
    month = jul,
    year = "2023",
    address = "Toronto, Canada",
    publisher = "Association for Computational Linguistics",
    url = "https://aclanthology.org/2023.acl-long.284/",
    doi = "10.18653/v1/2023.acl-long.284",
    pages = "5184--5207"
}

@article{tokenizercompression2024,
  title   = {Unpacking Tokenization: Evaluating Text Compression and its Correlation with Model Performance},
  author  = {{Goldman et al.}},
  journal = {arXiv preprint arXiv:2403.06265},
  year    = {2024},
  url     = {https://arxiv.org/abs/2403.06265}
}

@article{tokenizeropt2024,
  title   = {Getting the Most Out of Your Tokenizer for Pre-Training and Domain Adaptation},
  author  = {{Dagan et al.}},
  journal = {arXiv preprint arXiv:2402.01035},
  year    = {2024},
  url     = {https://arxiv.org/abs/2402.01035}
}

@article{towardTokenizationTheory2024,
  title   = {Toward a Theory of Tokenization in LLMs},
  author  = {Rajaraman, Nived and Jiao, Jiantao and Ramchandran, Kannan},
  journal = {arXiv preprint arXiv:2404.08335},
  year    = {2024},
  url     = {https://arxiv.org/abs/2404.08335}
}

@inproceedings{sennrich2016neural,
  author    = {Sennrich, Rico and Haddow, Barry and Birch, Alexandra},
  title     = {Neural Machine Translation of Rare Words with Subword Units},
  booktitle = {Proceedings of the 54th Annual Meeting of the Association for
               Computational Linguistics (Volume 1: Long Papers)},
  pages     = {1715--1725},
  year      = {2016},
  address   = {Berlin, Germany}
}

@inproceedings{kudo2018subword,
  author    = {Kudo, Taku},
  title     = {Subword Regularization: Improving Neural Network Translation
               Models with Multiple Subword Candidates},
  booktitle = {Proceedings of the 56th Annual Meeting of the Association for
               Computational Linguistics (Volume 1: Long Papers)},
  pages     = {66--75},
  year      = {2018},
  address   = {Melbourne, Australia}
}

@inproceedings{kudo2018sentencepiece,
  author    = {Kudo, Taku and Richardson, John},
  title     = {SentencePiece: A Simple and Language Independent Subword
               Tokenizer and Detokenizer for Neural Text Processing},
  booktitle = {Proceedings of the 2018 Conference on Empirical Methods in
               Natural Language Processing: System Demonstrations},
  pages     = {66--71},
  year      = {2018},
  address   = {Brussels, Belgium}
}

@article{wu2016gnmt,
  author  = {Wu, Yonghui and Schuster, Mike and Chen, Zhifeng and Le, Quoc V.
             and Norouzi, Mohammad and Macherey, Wolfgang and Krikun, Maxim
             and Cao, Yuan and Gao, Qin and Macherey, Klaus and others},
  title   = {Google's Neural Machine Translation System: Bridging the Gap
             Between Human and Machine Translation},
  journal = {arXiv preprint arXiv:1609.08144},
  year    = {2016}
}

@article{johnson2017google,
  author  = {Johnson, Melvin and Schuster, Mike and Le, Quoc V. and Krikun,
             Maxim and Wu, Yonghui and Chen, Zhifeng and Thorat, Nikhil and
             Vi{\'e}gas, Fernanda and Wattenberg, Martin and Corrado, Greg and
             Hughes, Macduff and Dean, Jeffrey},
  title   = {Google's Multilingual Neural Machine Translation System:
             Enabling Zero-Shot Translation},
  journal = {Transactions of the Association for Computational Linguistics},
  volume  = {5},
  pages   = {339--351},
  year    = {2017}
}

@article{radford2019gpt2,
  title     = {Language Models are Unsupervised Multitask Learners},
  author    = {Radford, Alec and Wu, Jeffrey and Child, Rewon and Luan, David and Amodei, Dario and Sutskever, Ilya},
  year      = {2019},
  journal   = {OpenAI Technical Report},
  note      = {Tech. rep.}
}

@inproceedings{mikolov2013distributed,
  author    = {Mikolov, Tomas and Sutskever, Ilya and Chen, Kai
               and Corrado, Greg S. and Dean, Jeffrey},
  title     = {Distributed Representations of Words and Phrases
               and Their Compositionality},
  booktitle = {Advances in Neural Information Processing Systems 26},
  pages     = {3111--3119},
  year      = {2013}
}

@inproceedings{chen1996,
author = {Chen, Stanley F. and Goodman, Joshua},
title = {An empirical study of smoothing techniques for language modeling},
year = {1996},
publisher = {Association for Computational Linguistics},
address = {USA},
url = {https://doi.org/10.3115/981863.981904},
doi = {10.3115/981863.981904},
booktitle = {Proceedings of the 34th Annual Meeting on Association for Computational Linguistics},
pages = {310–318},
numpages = {9},
location = {Santa Cruz, California},
series = {ACL '96}
}

@inproceedings{sutskever2014seq2seq,
  author    = {Sutskever, Ilya and Vinyals, Oriol and Le, Quoc V.},
  title     = {Sequence to Sequence Learning with Neural Networks},
  booktitle = {Advances in Neural Information Processing Systems 27},
  pages     = {3104--3112},
  year      = {2014}
}

@article{rissanen1984universal,
  title   = {Universal Coding, Information, Prediction, and Estimation},
  author  = {Rissanen, Jorma},
  journal = {IEEE Transactions on Information Theory},
  volume  = {30},
  number  = {4},
  pages   = {629--636},
  year    = {1984}
}

@article{xie1997minimax,
  title   = {Minimax Redundancy for the Class of Memoryless Sources},
  author  = {Xie, Qun and Barron, Andrew R.},
  journal = {IEEE Transactions on Information Theory},
  volume  = {43},
  number  = {2},
  pages   = {646--657},
  year    = {1997}
}

@article{drmota2003precise,
  title   = {Precise Minimax Redundancy and Regret},
  author  = {Drmota, Michael and Szpankowski, Wojciech},
  journal = {IEEE Transactions on Information Theory},
  volume  = {50},
  number  = {11},
  pages   = {2686--2707},
  year    = {2004}
}

@article{feder2025information,
  title   = {Information-Theoretic Framework for Understanding Modern Machine-Learning},
  author  = {Feder, Meir and Urbanke, Ruediger and Fogel, Yaniv},
  journal = {arXiv preprint arXiv:2506.07661},
  year    = {2025}
}

@inproceedings{schuster2012voice,
  author    = {Schuster, Mike and Nakajima, Kaisuke},
  title     = {Japanese and Korean Voice Search},
  booktitle = {Proceedings of the 2012 IEEE International Conference on
               Acoustics, Speech and Signal Processing (ICASSP)},
  pages     = {5149--5152},
  year      = {2012}
}

@inproceedings{bahdanau2015nmt,
  author    = {Bahdanau, Dzmitry and Cho, Kyunghyun and Bengio, Yoshua},
  title     = {Neural Machine Translation by Jointly Learning
               to Align and Translate},
  booktitle = {International Conference on Learning Representations},
  year      = {2015}
}

@inproceedings{brown2020gpt3,
  title     = {Language Models are Few-Shot Learners},
  author    = {Brown, Tom B. and Mann, Benjamin and Ryder, Nick and Subbiah,
               Melanie and Kaplan, Jared and Dhariwal, Prafulla et. al},
  booktitle = {Advances in Neural Information Processing Systems},
  volume    = {33},
  pages     = {1877--1901},
  year      = {2020}
}

@article{gage1994new,
  title={A new algorithm for data compression},
  author={Gage, Philip},
  journal={The C Users Journal},
  volume={12},
  number={2},
  pages={23--38},
  year={1994},
  publisher={R \& D Publications, Inc. Lawrence, KS, USA}
}

@ARTICLE{lz77,
  author={Ziv, J. and Lempel, A.},
  journal={IEEE Transactions on Information Theory}, 
  title={A universal algorithm for sequential data compression}, 
  year={1977},
  volume={23},
  number={3},
  pages={337-343},
  doi={10.1109/TIT.1977.1055714}}

@article{shani2025tokens,
  title={From tokens to thoughts: How LLMs and humans trade compression for meaning},
  author={Shani, Chen and Soffer, Liron and Jurafsky, Dan and LeCun, Yann and Shwartz-Ziv, Ravid},
  journal={arXiv preprint arXiv:2505.17117},
  year={2025}
}

@article{tishby2000information,
  title={The information bottleneck method},
  author={Tishby, Naftali and Pereira, Fernando C and Bialek, William},
  journal={arXiv preprint physics/0004057},
  year={2000}
}

@techreport{collet2018zstandard,
  title={Zstandard compression and the application/zstd media type},
  author={Collet, Yann and Kucherawy, Murray},
  year={2018}
}

@inproceedings{song2021fast,
  title={Fast wordpiece tokenization},
  author={Song, Xinying and Salcianu, Alex and Song, Yang and Dopson, Dave and Zhou, Denny},
  booktitle={Proceedings of the 2021 conference on empirical methods in natural language processing},
  pages={2089--2103},
  year={2021}
}

@article{raffel2020exploring,
  title={Exploring the limits of transfer learning with a unified text-to-text transformer},
  author={Raffel, Colin and Shazeer, Noam and Roberts, Adam and Lee, Katherine and Narang, Sharan and Matena, Michael and Zhou, Yanqi and Li, Wei and Liu, Peter J},
  journal={Journal of machine learning research},
  volume={21},
  number={140},
  pages={1--67},
  year={2020}
}

@inproceedings{ortiz-suarez-oscar,
    title = "A Monolingual Approach to Contextualized Word Embeddings for Mid-Resource Languages",
    author = "Ortiz Su{'a}rez, Pedro Javier  and
      Romary, Laurent  and
      Sagot, Benoit",
    booktitle = "Proceedings of the 58th Annual Meeting of the Association for Computational Linguistics",
    month = jul,
    year = "2020",
    address = "Online",
    publisher = "Association for Computational Linguistics",
    url = "https://www.aclweb.org/anthology/2020.acl-main.156",
    pages = "1703--1714",
}

@article{li2023starcoder,
  title={Starcoder: may the source be with you!},
  author={Li, Raymond and Allal, Loubna Ben and Zi, Yangtian and Muennighoff, Niklas and Kocetkov, Denis and Mou, Chenghao and Marone, Marc and Akiki, Christopher and Li, Jia and Chim, Jenny and others},
  journal={arXiv preprint arXiv:2305.06161},
  year={2023}
}

@article{cobbe2021gsm8k,
  title={Training Verifiers to Solve Math Word Problems},
  author={Cobbe, Karl and Kosaraju, Vineet and Bavarian, Mohammad and Chen, Mark and Jun, Heewoo and Kaiser, Lukasz and Plappert, Matthias and Tworek, Jerry and Hilton, Jacob and Nakano, Reiichiro and Hesse, Christopher and Schulman, John},
  journal={arXiv preprint arXiv:2110.14168},
  year={2021}
}

@misc{CodeParrot, title={Codeparrot/codeparrot-clean · datasets at hugging face}, url={https://huggingface.co/datasets/codeparrot/codeparrot-clean}, journal={codeparrot/codeparrot-clean · Datasets at Hugging Face}, author={CodeParrot}}

@article{xlm-roberta,
  author    = {Alexis Conneau and
               Kartikay Khandelwal and
               Naman Goyal and
               Vishrav Chaudhary and
               Guillaume Wenzek and
               Francisco Guzm{\'{a}}n and
               Edouard Grave and
               Myle Ott and
               Luke Zettlemoyer and
               Veselin Stoyanov},
  title     = {Unsupervised Cross-lingual Representation Learning at Scale},
  journal   = {CoRR},
  volume    = {abs/1911.02116},
  year      = {2019},
  url       = {http://arxiv.org/abs/1911.02116},
  eprinttype = {arXiv},
  eprint    = {1911.02116},
  bibsource = {dblp computer science bibliography, https://dblp.org}
}

@software{tiktoken,
  author = {{OpenAI}},
  title = {tiktoken},
  url = {https://github.com/openai/tiktoken},
  version = {0.9.0},
  year = {2025},
}

@article{achiam2023gpt,
  title={Gpt-4 technical report},
  author = {{OpenAI}},
  journal={arXiv preprint arXiv:2303.08774},
  year={2023}
}
}

\newpage
\onecolumn

\appendix

\subsection{Correlation between compression and tokenization}
\label{app-A}

For each domain and tokenizer $T$, we draw $n = 1000$ documents
$\{d_i\}_{i=1}^n$.  
On each document we compute:
\[
x_i = f(d_i), \text{ } y_i = g_T(d_i),
\]
where $f$ denotes the intrinsic compressibility of the document
(e.g., zstd bits per character), and $g_T$ denotes the tokenization density under tokenizer $T$ (tokens per character).
We then report the Pearson correlation:
\[
r_T = 
\frac{\sum_i (x_i - \bar{x})(y_i - \bar{y})}
     {\sqrt{\sum_i (x_i - \bar{x})^2}\,
      \sqrt{\sum_i (y_i - \bar{y})^2}},
\]
where $\bar{x}$ and $\bar{y}$ are the sample means.

This measures whether documents that are intrinsically harder to compress (higher $x_i$) also receive a finer tokenization under $T$ (higher $y_i$). We also computed Spearman rank correlations and observed the same qualitative patterns, so for brevity we only report Pearson $r_T$ in the table below:

\begin{table}[h]
\centering
\begin{tabular}{|l|r|r|r|r|}
\hline
\textbf{Domain} & \texttt{o200k\_base} & \texttt{cl100k\_base} & \texttt{p50k\_base} & \texttt{gpt2} \\
\hline
News & $0.329$ & $0.326$ & $0.347$ & $0.337$ \\
Code & $-0.109$ & $-0.094$ & $-0.164$ & $-0.510$ \\
Math & $0.435$ & $0.407$ & $0.311$ & $0.311$ \\
\hline
\end{tabular}
\vspace{2pt}
\caption{Correlation between tokenization and compression for each (domain, tokenizer)}
\label{tab:compression-corr}
\end{table}

On news and math, Pearson correlations of roughly $0.3$--$0.4$ indicate that documents which are intrinsically harder to compress also receive a finer tokenization, so GPT token counts track \texttt{zstd} complexity reasonably well. In contrast, for code the \texttt{gpt2} tokenizer shows a strong negative correlation ($r \approx -0.5$), meaning it uses \emph{more} tokens precisely on snippets that\texttt{zstd}zstd finds easiest, while the newer, code-aware tokenizers (\texttt{cl100k\_base}, \texttt{o200k\_base}) largely mitigate this mismatch
(correlations near zero).

\subsection{Tokenizer Descriptions from Section~\ref{sec:learned}:}
\label{app:tokenizers}

\textbf{Byte-Pair Encoding (BPE):} BPE extends the classic compression algorithm of Gage~\cite{gage1994new} to
subword segmentation by iteratively merging frequent symbol pairs until a target vocabulary size is reached. The subword variant for various NLP applications was popularized in~\cite{sennrich2016neural} and is now one of the dominant choices for pretraining large language models, including the GPT family of models~\cite{radford2019gpt2,brown2020gpt3}.\footnote{In GPT-style architectures, BPE is typically implemented at the byte level, so tokenization operates directly on raw bytes rather than characters or words, yielding a deterministic and reversible mapping between text and tokens. This byte-level formulation supports arbitrary Unicode and non-linguistic inputs such as code, markup, and logs without out-of-vocabulary issues, which is crucial for robust, stable training at web scale~\cite{radford2019gpt2,brown2020gpt3}.
}

\textbf{Unigram:}
The unigram tokenizer models a vocabulary of subword units with a discrete unigram LM and performs segmentation by (approximately) maximizing the likelihood of the observed text under this model~\cite{kudo2018subword,kudo2018sentencepiece}. Unlike BPE’s greedy merge process, the unigram approach tends to produce subwords that align more closely with morphology and is shown to be superior to BPE in some downstream tasks.

\textbf{WordPiece:} WordPiece is a data-driven subword vocabulary construction method originally developed for Japanese \& Korean voice search~\cite{schuster2012voice} and later widely adopted in Google’s machine translation models~\cite{wu2016gnmt,johnson2017google}. It builds a subword inventory that balances coverage of frequent words while decomposing rare words into meaningful pieces.

\textbf{WordLevel:} As a simpler baseline, we include a word-level tokenizer that treats whitespace-delimited tokens (plus a set of special symbols) as atomic
units. This corresponds to the the traditional word-based representation used in classic $n$-gram LMs~\cite{chen1996} and early neural LMs and machine translation tasks~\cite{mikolov2013distributed,sutskever2014seq2seq,bahdanau2015nmt} before the widespread adoption of subword methods.

\subsection{Trade-off between Compression Ratio and Vocabulary Size:}
\label{app:modelcapacity}
For a fixed tokenizer $T$ with vocabulary size $\lvert V\rvert = K$, the expected code length on a test distribution $P$ can be decomposed as follows:
\[
\mathbb{E}[\ell] \approx H(P_T) + R_n,
\]
where $H(P_T)$ is the cross-entropy of the token distribution induced by $P$ under $T$ (a model--mismatch/capacity term) and $R_n$ denotes the redundancy term (an estimation error term) for a universal i.i.d.\ token coder trained on $n$ characters of corpus. 

(\emph{Sketch:} For a universal code with coding distribution $Q_n$ on token sequences, the expected length satisfies $\tfrac{1}{n}\mathbb{E}[\ell_n]
= H(P_T) + \tfrac{1}{n}D(P_T^{\otimes n} \Vert Q_n)$; the second term is the redundancy $R_n$ up to normalization.)

From standard universal coding bounds, we have:
\[
R_n \;\sim\; \frac{\lvert V\rvert - 1}{2n} \log n.
\]
up to $o(1/n)$ terms and an additive $O(1/n)$ constant
(see,~\cite{rissanen1984universal,xie1997minimax,drmota2003precise}). So, as the training data ($n$) increases, the estimation error has a monotone decrease.

(\emph{Sketch:} The family of i.i.d.\ distributions over a $K$-symbol alphabet has parameter dimension $d = K-1$. The minimax redundancy for such a $d$-dimensional parametric family scales as $\tfrac{d}{2}\log n + O(1)$ in total, so the per-symbol redundancy behaves like $\tfrac{d}{2n}\log n$.)

Let $T_n$ denote the tokenizer learned from $n$ characters of training corpus, and
write $P_{T_n}$ for the induced token distribution on the test source. Then:
\begin{itemize}
    \item When $K$ is large, the tokenizer learned from more data
    converges (in an empirical-risk sense) to a dictionary
    $T^\star$ that is near-optimal for both train and test distributions, so $H(P_{T_n})$ stays essentially flat while
    $R_n$ decreases with $n$, giving the monotone improvement we observe.
    
    (\emph{Sketch:} As $n\to\infty$, empirical token frequencies under the training distribution $Q$ converge to their expectations, so any reasonable training criterion approximates the expected cross-entropy under $Q$, and $T_n$ drifts toward a fixed optimizer $T^\star$; the universal-coding term $R_n$ then decays like $\tfrac{K-1}{2n}\log n$.)

    \item When $K$ is small, the optimal dictionary $T^\star_Q$ for the large training corpus $Q$ can be far from the dictionary that would minimize $H(P_T)$ for the test file; as $n$ grows the learned $T_n$ moves toward $T^\star_Q$, so
    $H(P_{T_n})$ can increase (worse segmentation for the test distribution) even though $R_n$ shrinks, yielding the non-monotone behavior in~\autoref{fig:compression-grid16k}.
    
    (\emph{Sketch:} With a tight $K$-token constraint, the training-optimized dictionary must compromise across many patterns present in $Q$; if the test distribution $P$ is a different mixture of these patterns, the $T^\star_Q$
    that minimizes cross-entropy on $Q$ need not minimize $H(P_T)$, so convergence of $T_n$ to $T^\star_Q$ can move $P_{T_n}$ \emph{away} from the $P$-optimal tokenizer.)
\end{itemize}

This provides a theoretical justification for our takeaway that more data can hurt when the model class is mis-specified and under-parameterized.

\subsection{More on Tokenization as a Channel}
\label{app:tok-chan}

So far, we have primarily investigated the efficacy of tokenizers primarily as \emph{compressors}: they map UTF-8 byte streams into shorter token sequences, where we evaluate them via compression proxies (bytes per token, LZ bits per character) and local predictability (token-level $k$-gram entropies). In this subsection, we adopt a complementary view, following~\cite{zouhar2023tokenization}, and regard tokenization as a \emph{noiseless discrete channel} that mediates between raw text and downstream models. Formally, a tokenizer $T$ induces a deterministic mapping $T: \mathcal{X}^* \to \mathcal{V}^*$, where $\mathcal{X}$ is the input alphabet (UTF-8 bytes or characters) and $\mathcal{V}$ is a finite token vocabulary of size $|\mathcal{V}| = K$. Once text is mapped to a token sequence $(t_1,\dots,t_n) \in \mathcal{V}^n$, this sequence is ``transmitted" over a noiseless $K$-ary channel to the language model. The channel has capacity $ C_{\text{token}} = \log_2 K \text{ } \text{bits per token}$, achieved when the token distribution is uniform. In contrast, the empirical token distribution induced by real corpora is significantly skewed; let $H_1(T; D)$ denote the unigram entropy of the token stream produced by $T$ on a test corpus $D$, as estimated in Sec.~V.
Then, only a fraction
\[
  \eta(T; D) \;\triangleq\; \frac{H_1(T; D)}{\log_2 K}
\]
of the available channel capacity is actually \emph{used} on average. Following~\cite{zouhar2023tokenization}, we also consider a Rényi analogue: for $\alpha>1$,
\[
\eta_\alpha(T;D) \;\triangleq\; \frac{H_\alpha(T;D)}{\log_2 K},
\]
which places greater weight on the head of the token distribution (e.g., $\alpha {=} 2$ corresponds to collision entropy). This channel view helps tie together several of our empirical observations and motivates concrete future directions.

First, as a function of tokenizer training size, we previously observed that the unigram entropy $H_1$ increases while higher-order conditional entropies $H_k$ for $k \ge 2$ decrease (Fig.~\ref{fig:kthorderEnglish}), indicating that the token distribution becomes less skewed and the token stream more locally predictable. Figure~\ref{fig:cap} revisits this through a channel lens by plotting the corresponding efficiencies for English tokenizers with $K = 16\text{k}$ as the training size grows from $10^3$ to $10^8$ characters. In terms of Shannon utilization, all tokenizers exhibit a rapid rise between $10^3$ and $10^5$ characters (from $\eta \approx 0.46$--$0.50$ up to $\eta \approx 0.64$--$0.73$), followed by clear saturation by $10^6$ characters and beyond (stabilizing around $\eta \approx 0.68$--$0.77$), suggesting that relatively modest tokenizer-training data already captures most of the attainable marginal ``channel usage'', with additional data yielding only diminishing returns. 

\noindent
Among subword schemes, BPE and WordPiece plateau near $\eta \approx 0.75$--$0.77$, whereas Unigram remains lower (around $\eta \approx 0.69$), consistent with a more skewed unigram distribution and hence more residual redundancy at the token level. While WordLevel eventually attains comparable Shannon utilization at large training sizes, Fig.~\ref{fig:kthorderEnglish} shows that it leaves substantially larger conditional entropies $H_k$ for $k\ge 2$, i.e., it uses the token alphabet more uniformly but absorbs less low-order structure into individual symbols.

Rényi efficiency reveals a complementary effect that is obscured by Shannon utilization alone. For $\alpha=2$, $\eta_\alpha$ is uniformly smaller than $\eta$ and, for BPE/WordPiece/Unigram, increases up to roughly $10^5$ characters before slightly declining thereafter. This divergence suggests that while additional tokenizer-training data continues to populate the long tail of rare tokens (raising $H_1$ and hence $\eta$), it can also concentrate probability mass into a smaller set of extremely frequent tokens, increasing collision probability and reducing $H_2$, precisely the kind of head concentration that Rényi-based criteria in~\cite{zouhar2023tokenization} are designed to penalize. WordLevel behaves differently: at very small training sizes its efficiency is extremely low (especially in Rényi), consistent with severe vocabulary undercoverage and a dominant \texttt{<unk>} mass, and even at larger sizes it exhibits substantially lower $\eta_2$ than subword tokenizers at comparable $\eta$, reflecting the heavy head of word-frequency distributions in English.

Overall, combining these efficiency trends with our earlier $H_k$ results indicates that subword tokenizers (notably BPE/WordPiece) offer the most favorable trade-off for modeling: they achieve high marginal channel utilization while absorbing substantial low-order structure (lower $H_k$) without inducing a concentration among the most frequent symbols as word-level vocabularies, consistent with their widespread use in modern LLM pipelines. These efficiency measures suggest concrete links to downstream modeling beyond compression. 
Under train--test mismatch (Sec.~VI), these same learned tokenizers can have very different effective utilization $\eta(T; D)$ and conditional entropies $H_k$. When we train on English but test on Turkish, Chinese, or code, the compression ratio no longer improves monotonically with training size and the higher-order entropies remain bounded away from zero (Fig.~\ref{fig:compression-domainmismatch}). From the channel perspective, we are now using a \emph{codebook} optimized for one source distribution to transmit a different source over the same $K$-ary channel: many tokens become rare or unused, and the model sees token streams that are both less compressive and harder to predict. This provides a unified way to interpret the brittleness we observe under distribution shift, and connects naturally to recent work that explicitly studies tokenization as a noiseless channel between meaning and surface form~\cite{zouhar2023tokenization}.

\subsection{Compression ratios across other domains}

\begin{figure*}[ht]
    \centering
    \vspace{-0.25cm}
    \subfigure[Turkish]{
        \includegraphics[width=0.3\linewidth]{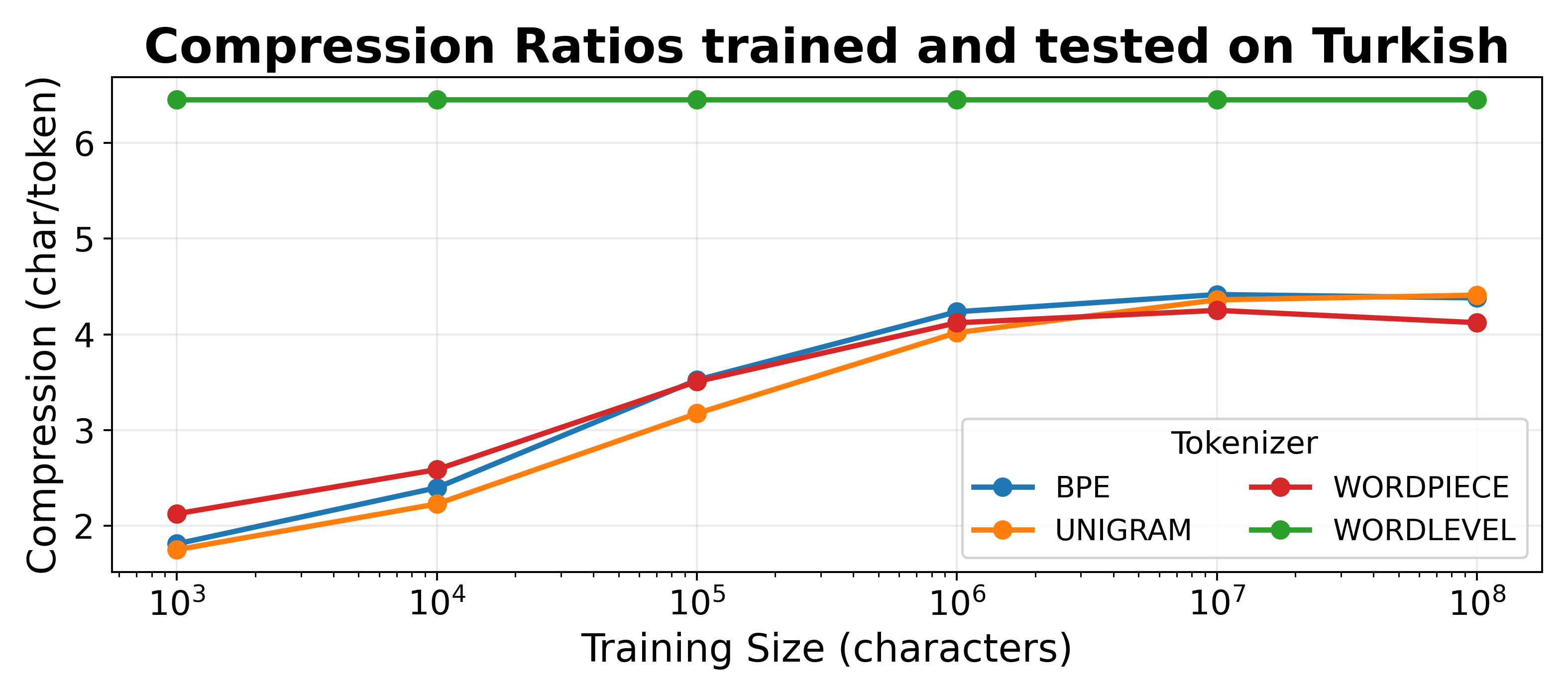}
    } \hspace{-0.45cm}
    \subfigure[Chinese]{
        \includegraphics[width=0.3\linewidth]{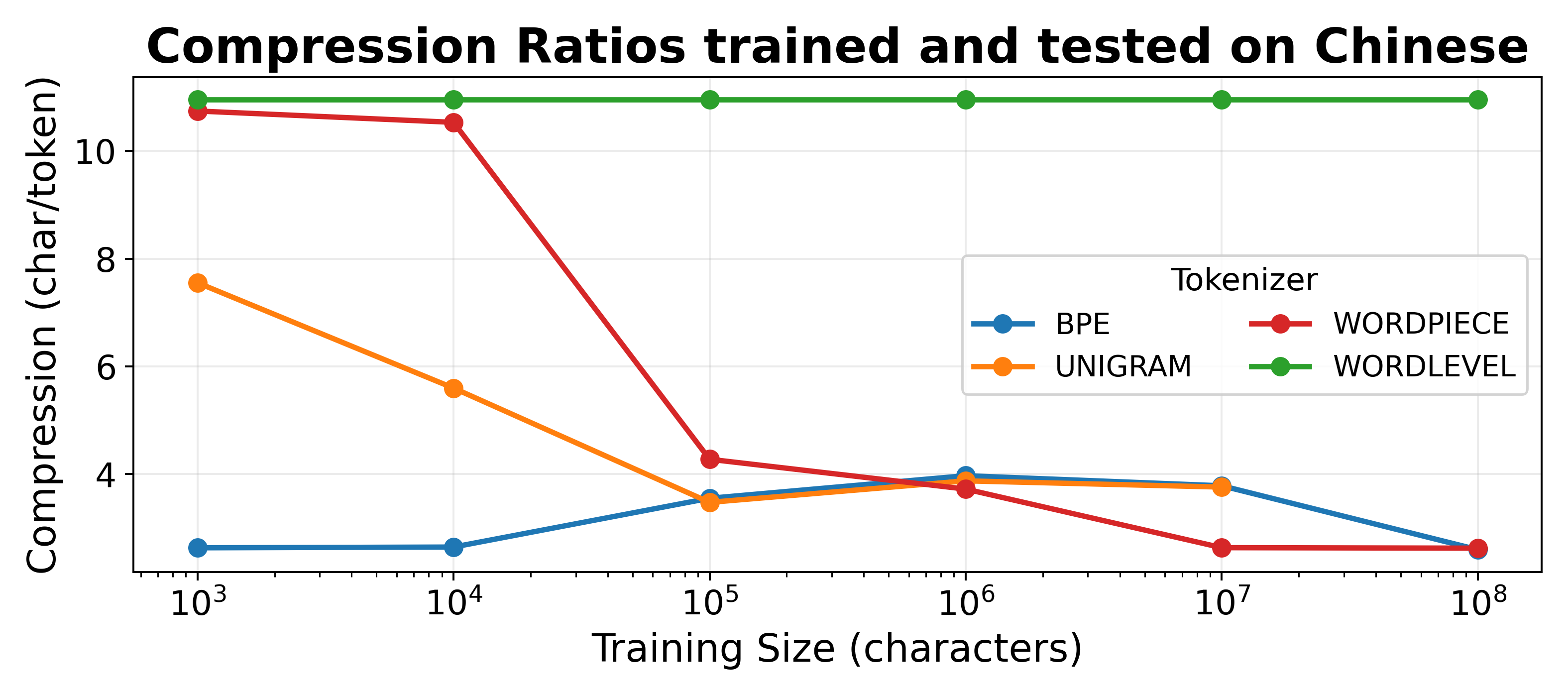}
    } \hspace{-0.45cm}
    \subfigure[Chinese-Latin]{
        \includegraphics[width=0.3\linewidth]{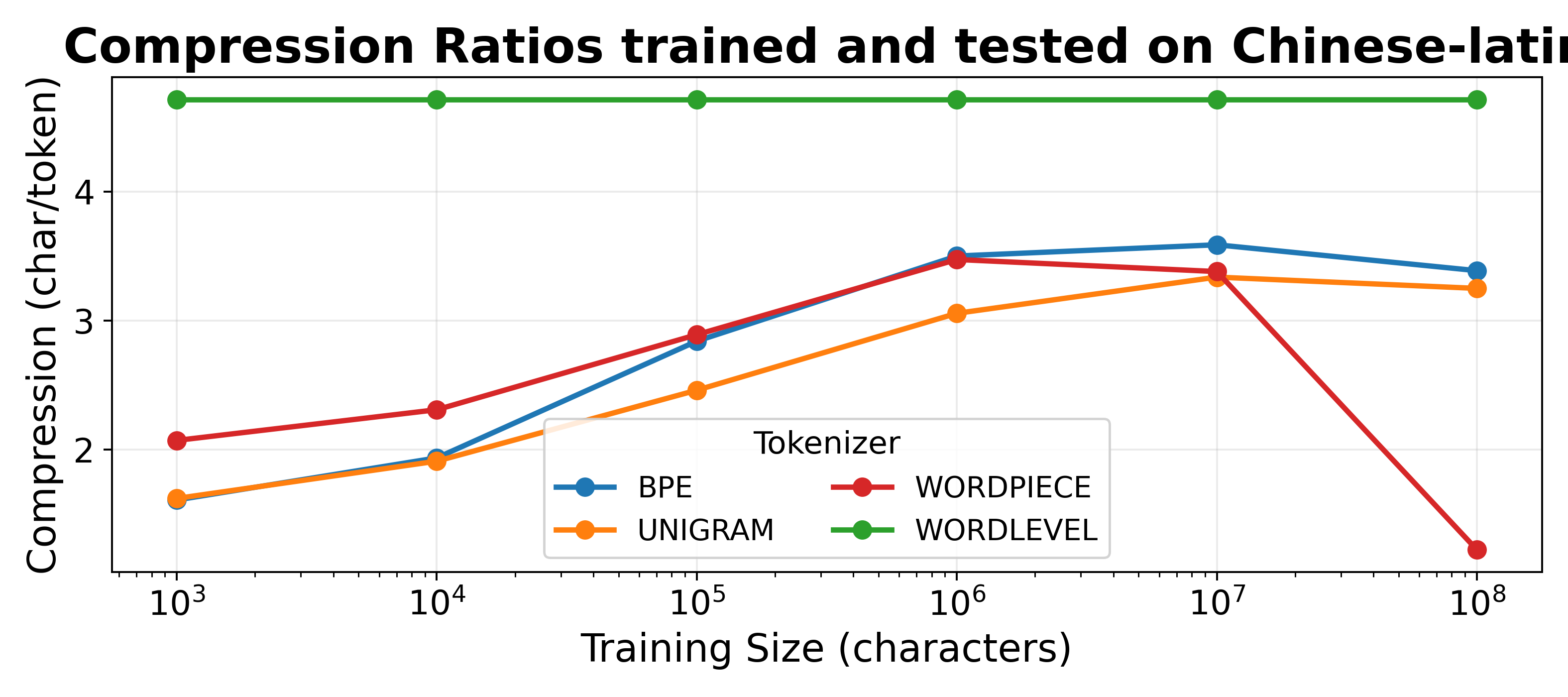}
    }
    \caption{Compression ratios across domains (vocabulary size = 16k)}
    \label{fig:compression-grid16k-app}
\end{figure*}

\begin{figure*}[ht]
    \centering
    \vspace{-0.25cm}
    \subfigure[Turkish]{
        \includegraphics[width=0.3\linewidth]{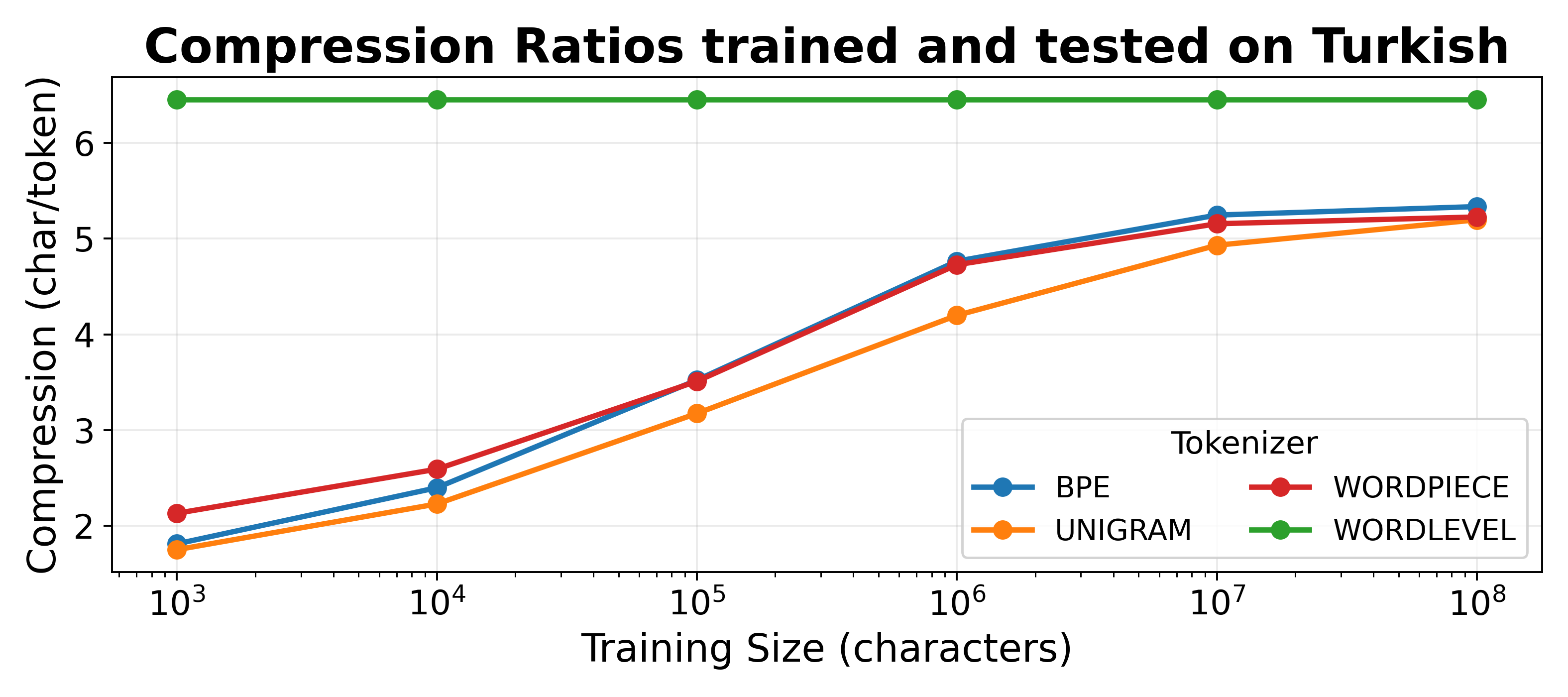}
    } \hspace{-0.45cm}
    \subfigure[Chinese]{
        \includegraphics[width=0.3\linewidth]{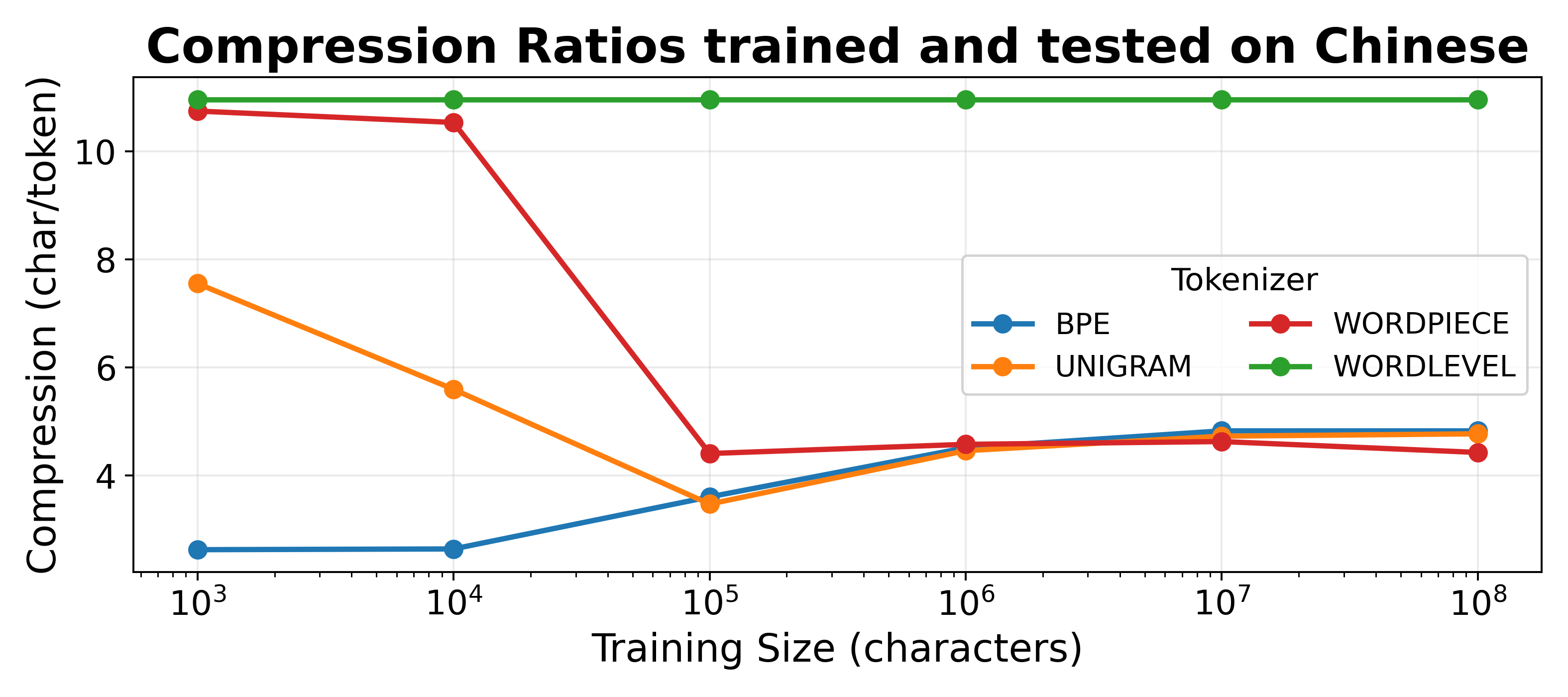}
    } \hspace{-0.45cm}
    \subfigure[Chinese-Latin]{
        \includegraphics[width=0.3\linewidth]{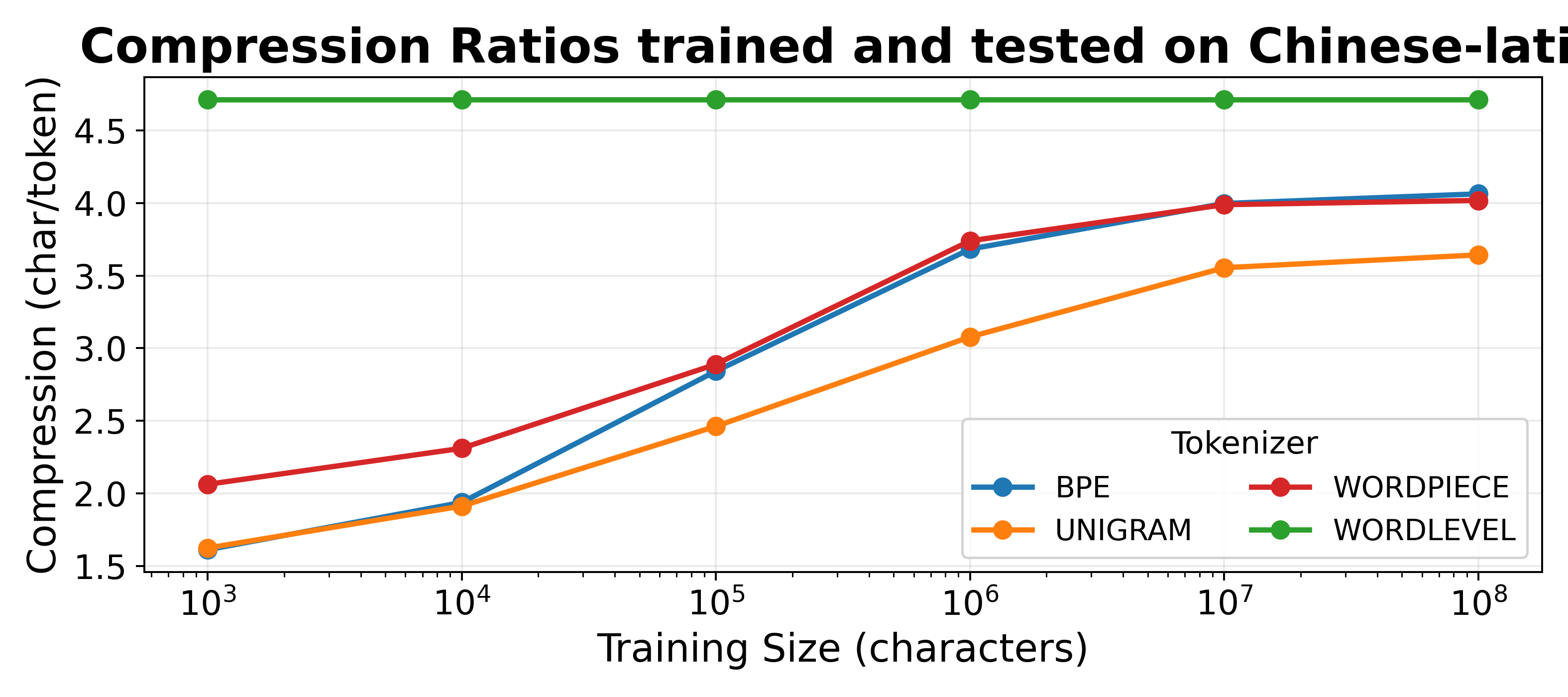}
    }
    \caption{Compression ratios across domains (vocabulary size = 64k)}
    \label{fig:compression-grid64k-app}
\end{figure*}

\begin{figure*}[ht]
    \centering
    \vspace{-0.25cm}
    \subfigure[Chinese]{
        \includegraphics[width=0.3\linewidth]{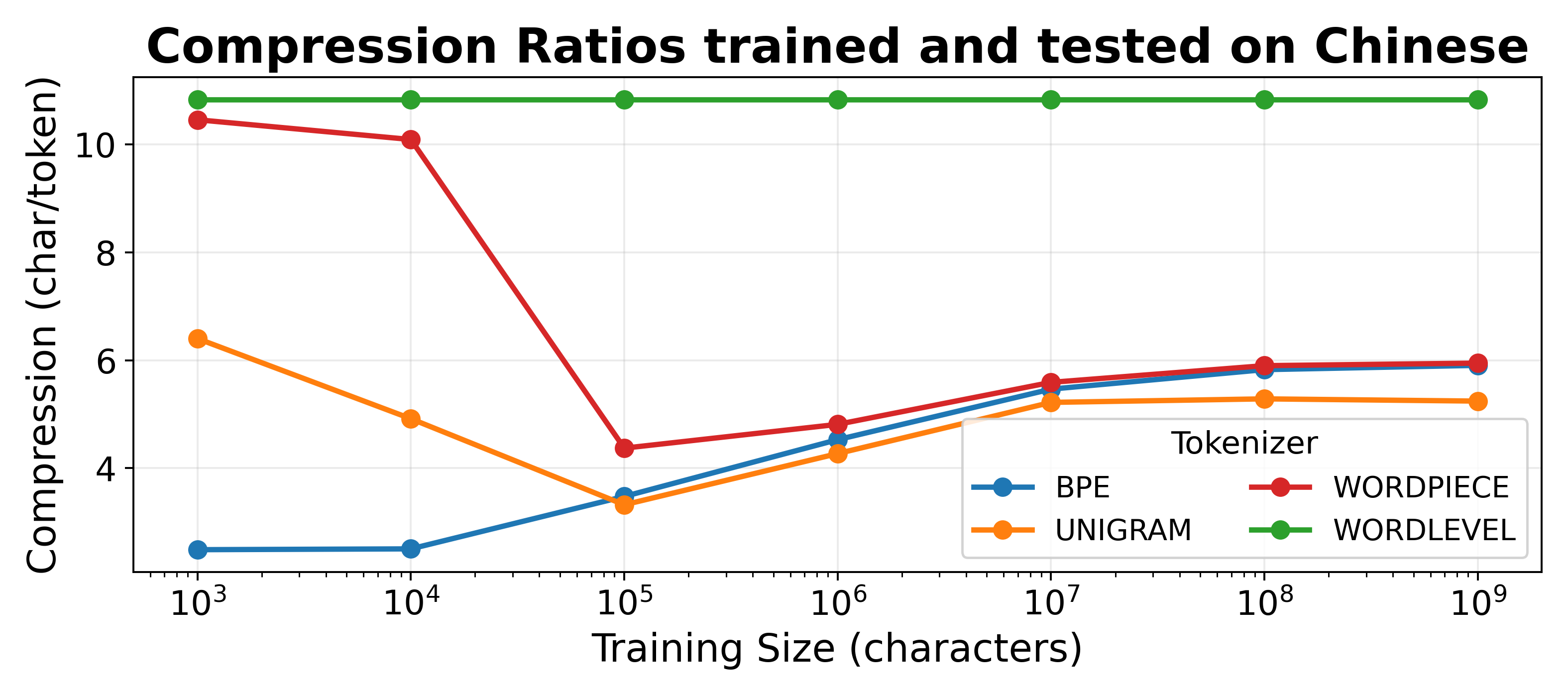}
    } \hspace{-0.45cm}
    \subfigure[Chinese-Latin]{
        \includegraphics[width=0.3\linewidth]{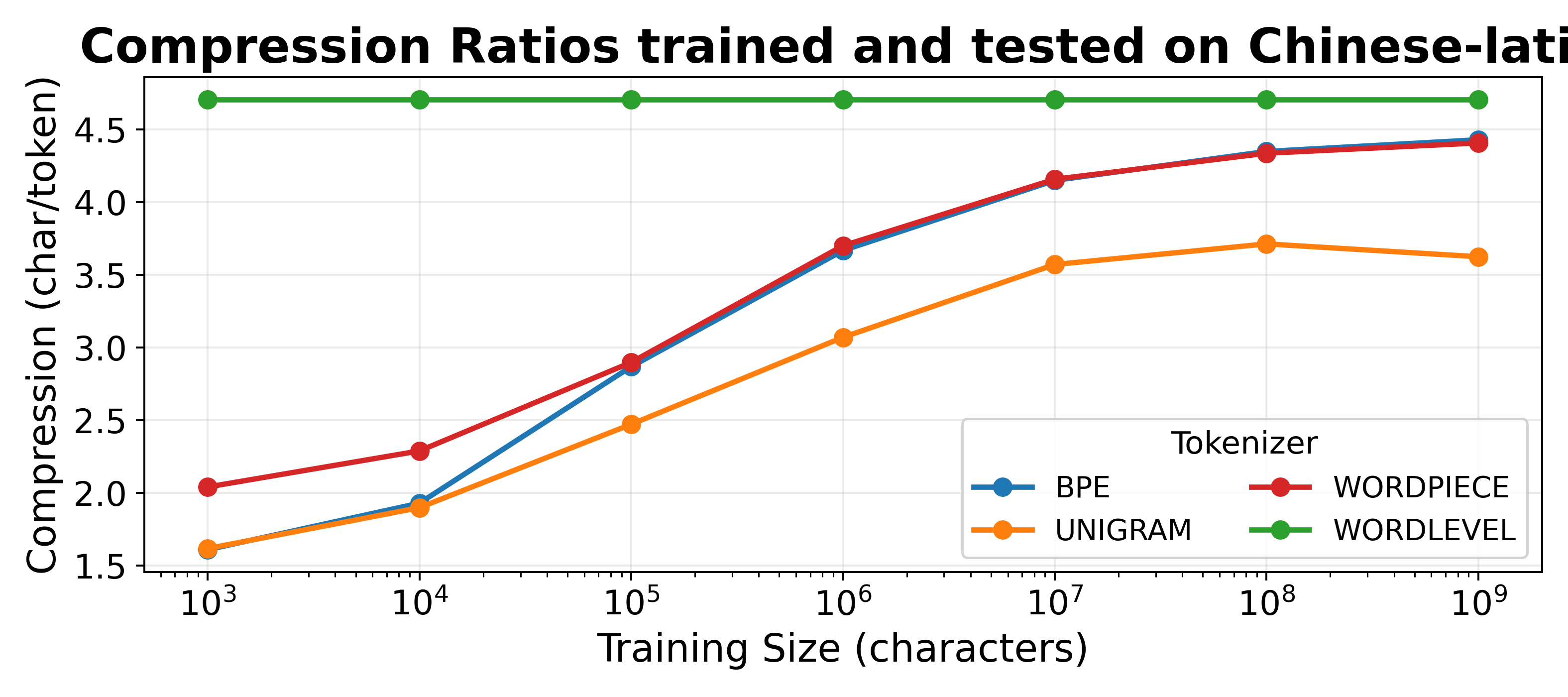}
    }
    \caption{Compression ratios across domains (vocabulary size = 500k)}
    \label{fig:compression-grid64k-app-2}
\end{figure*}

\begin{figure}[ht!]
    \centering
    \vspace{-0.25cm}
    \subfigure[Chinese]{
        \includegraphics[width=0.3\linewidth]{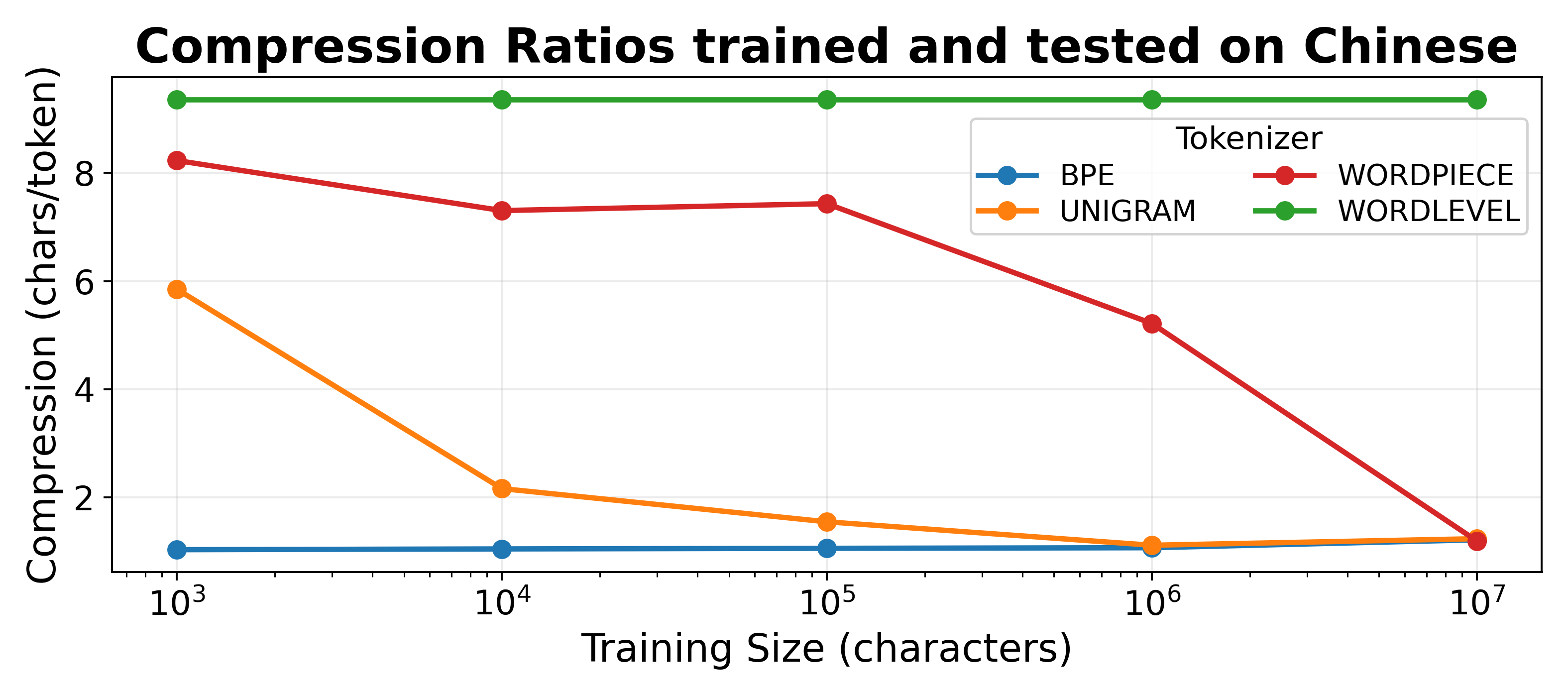}
    } \hspace{-0.45cm}
    \subfigure[Chinese-Latin]{
        \includegraphics[width=0.3\linewidth]{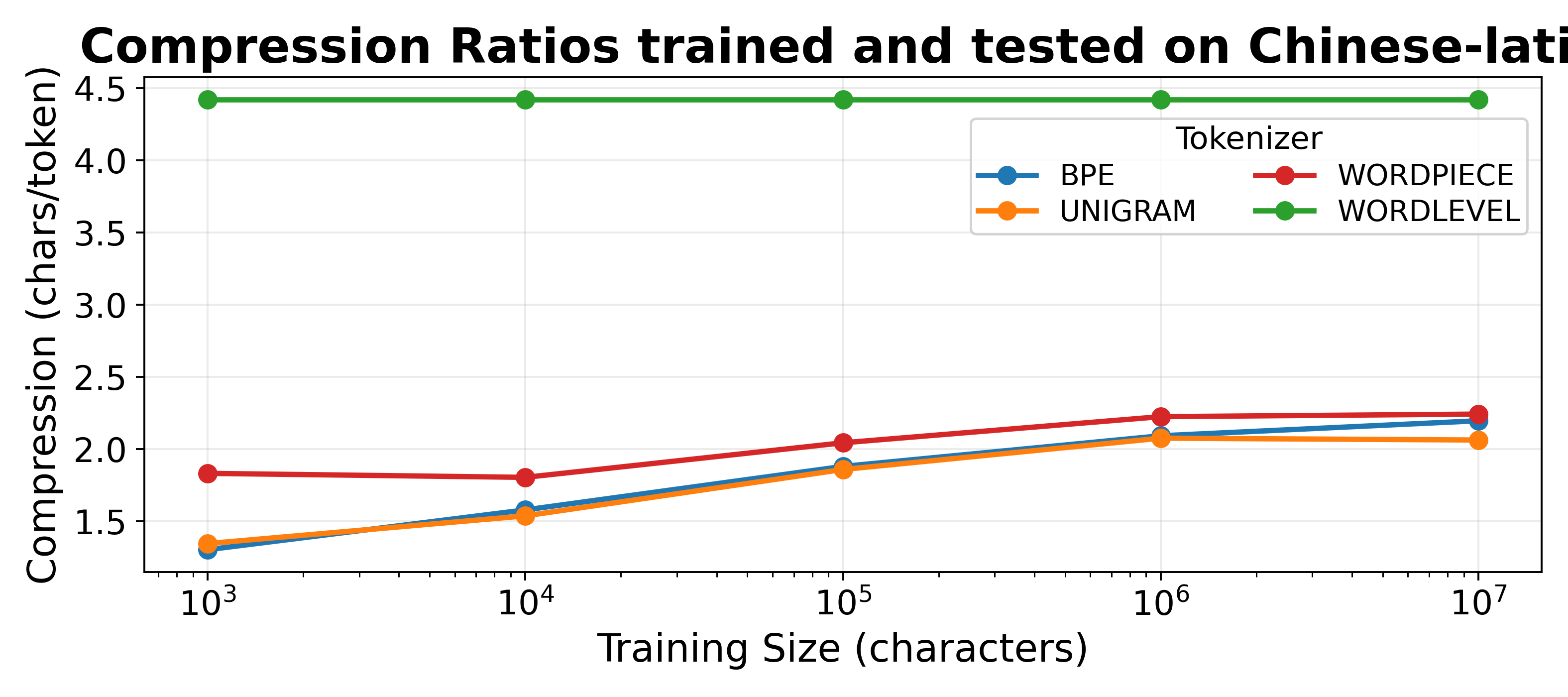}
    }
    \caption{Compression ratios in domain mismatch (vocabulary size = 16k).
    Trained on English, and tested on (a) Chinese (b) Chinese-Latin.}
    \label{fig:compression-domainmismatch2}
\end{figure}

\begin{figure*}[ht]
\centering
    \includegraphics[width=0.75\linewidth]{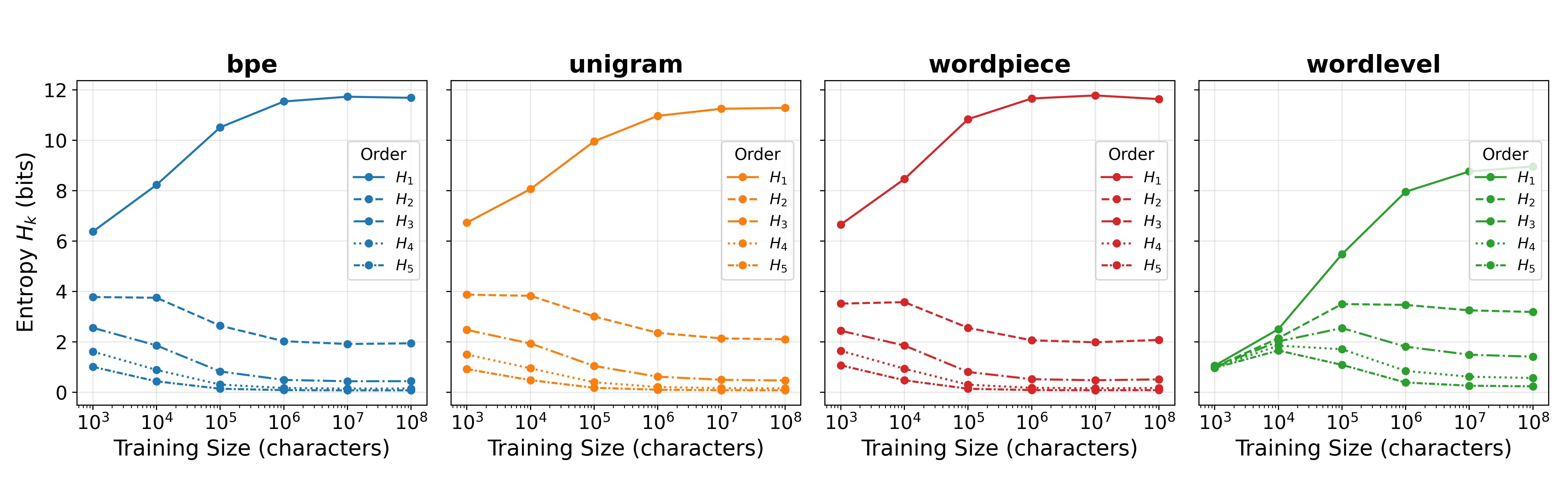}
    \caption{Tokenizer $k$-gram entropy results trained and tested on Turkish for vocabulary size 16k. \looseness=-2}
\label{fig:ent1}
\end{figure*}

\begin{figure*}[ht]
\centering
    \includegraphics[width=0.75\linewidth]{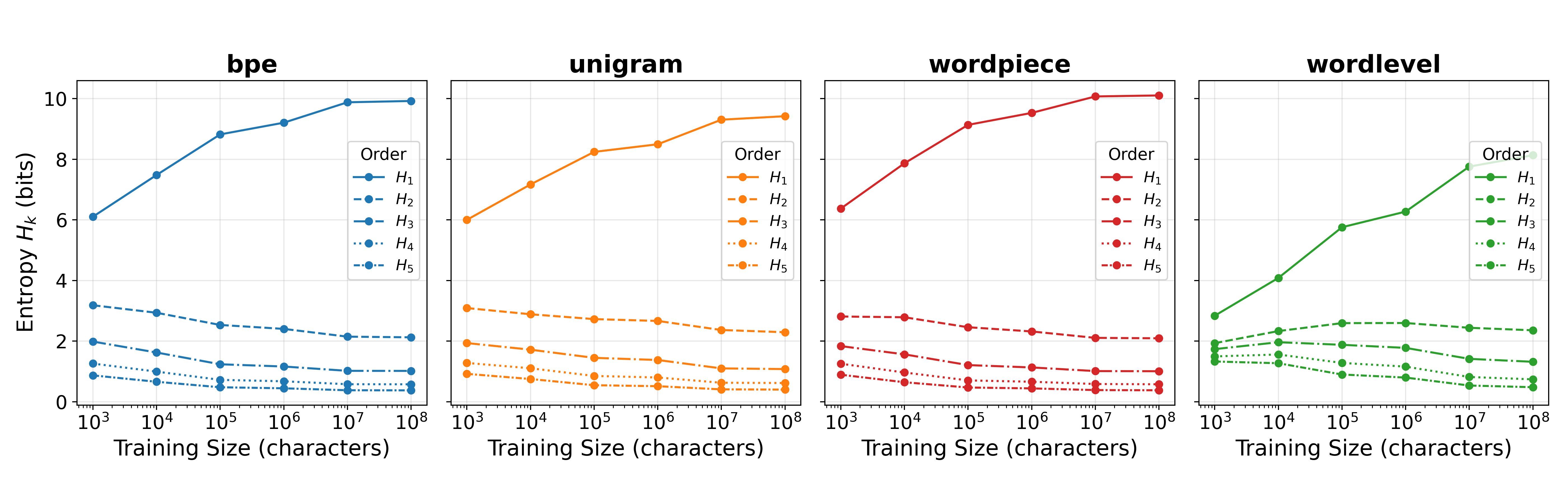}
    \caption{Tokenizer $k$-gram entropy results trained and tested on Code for vocabulary size 16k. \looseness=-2}
\label{fig:ent2}
\end{figure*}

\begin{figure*}[ht]
\centering
    \includegraphics[width=0.75\linewidth]{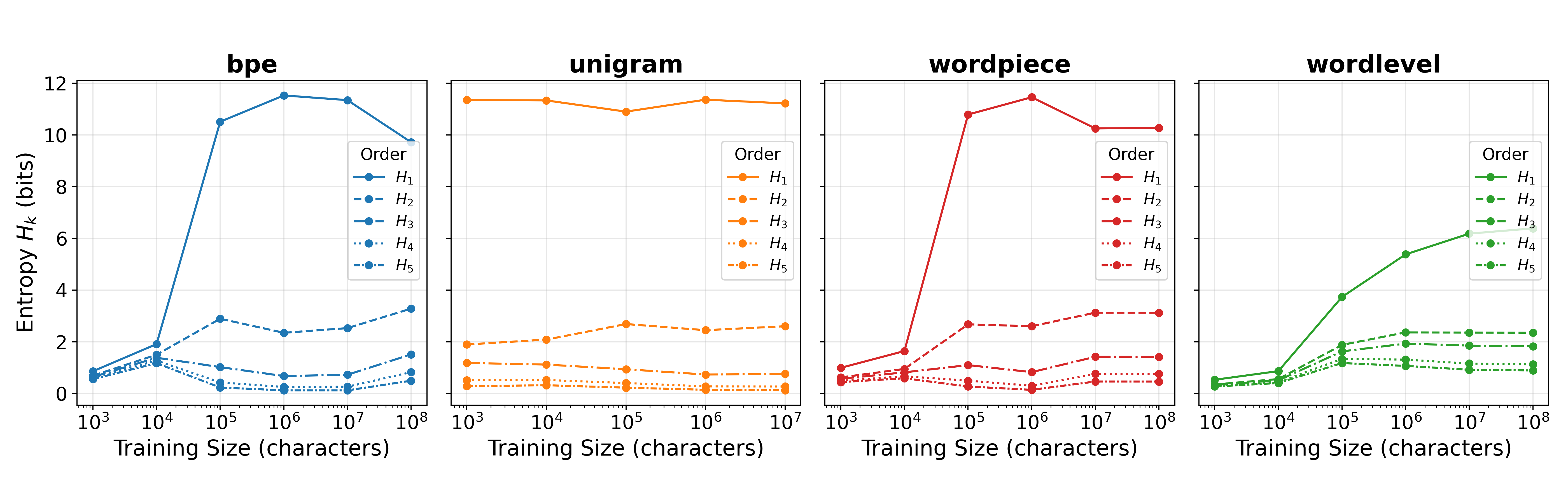}
    \caption{Tokenizer $k$-gram entropy results trained and tested on Chinese for vocabulary size 16k. \looseness=-2}
\label{fig:ent2x}
\end{figure*}

\begin{figure*}[ht]
\centering
    \includegraphics[width=0.75\linewidth]{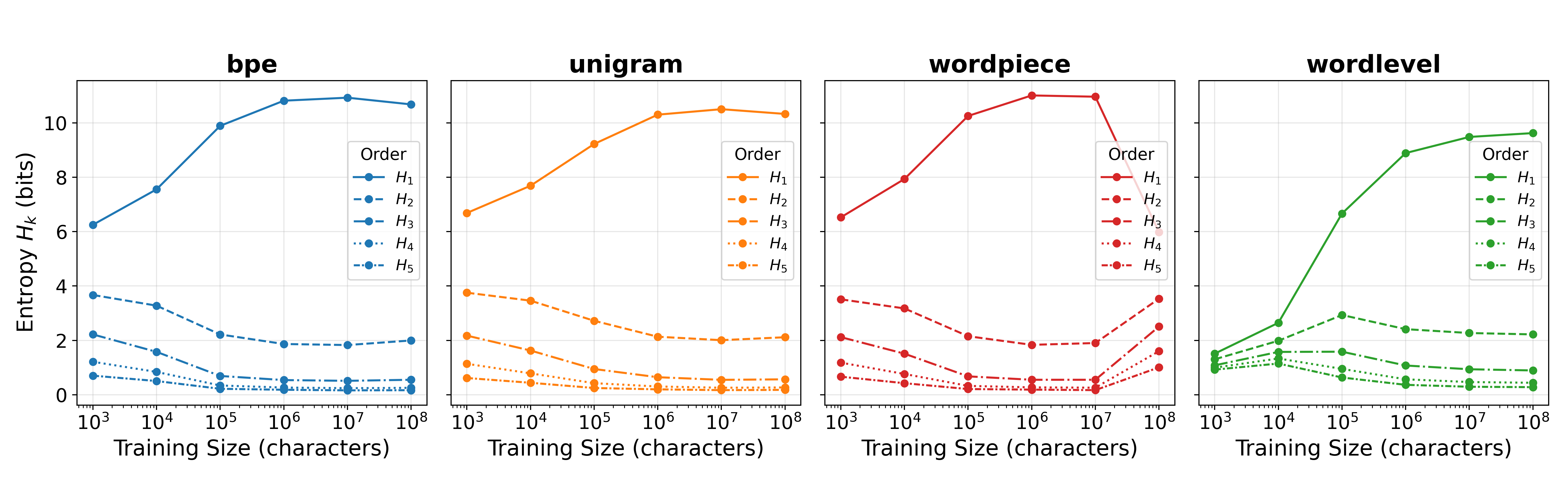}
    \caption{Tokenizer $k$-gram entropy results trained and tested on Chinese-Latin for vocabulary size 16k. \looseness=-2}
\label{fig:ent2y}
\end{figure*}

\begin{figure*}[ht]
\centering
    \includegraphics[width=0.75\linewidth]{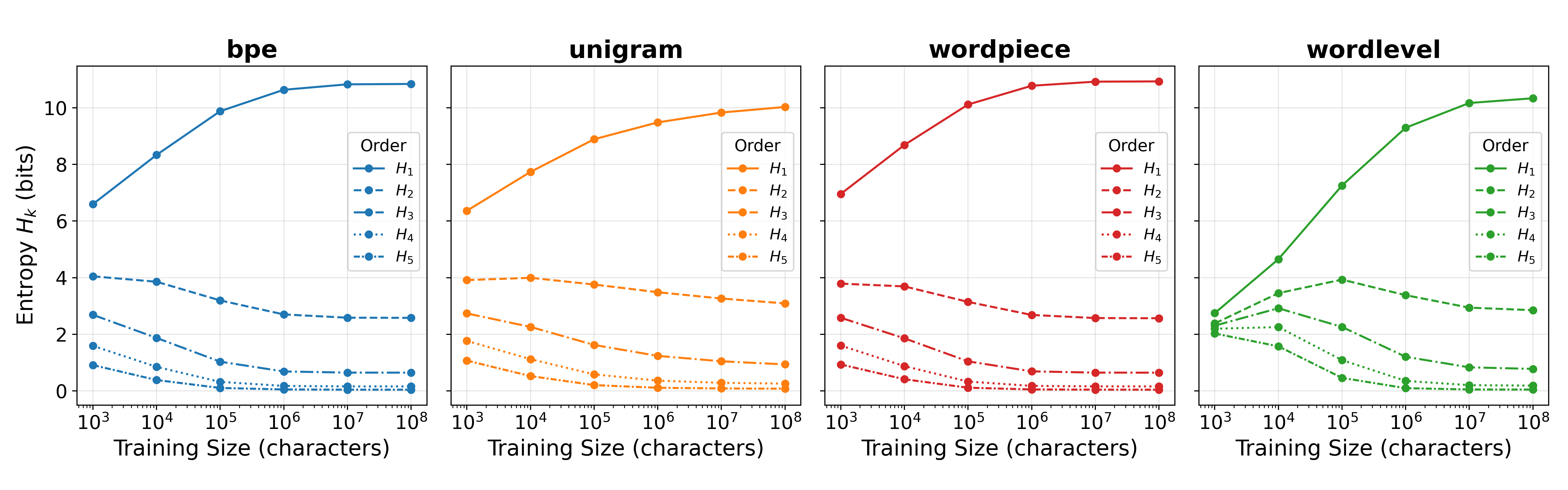}
    \caption{Tokenizer $k$-gram entropy results trained and tested on English for vocabulary size 64k. \looseness=-2}
\label{fig:ent3}
\end{figure*}

\begin{figure*}[ht]
\centering
    \includegraphics[width=0.75\linewidth]{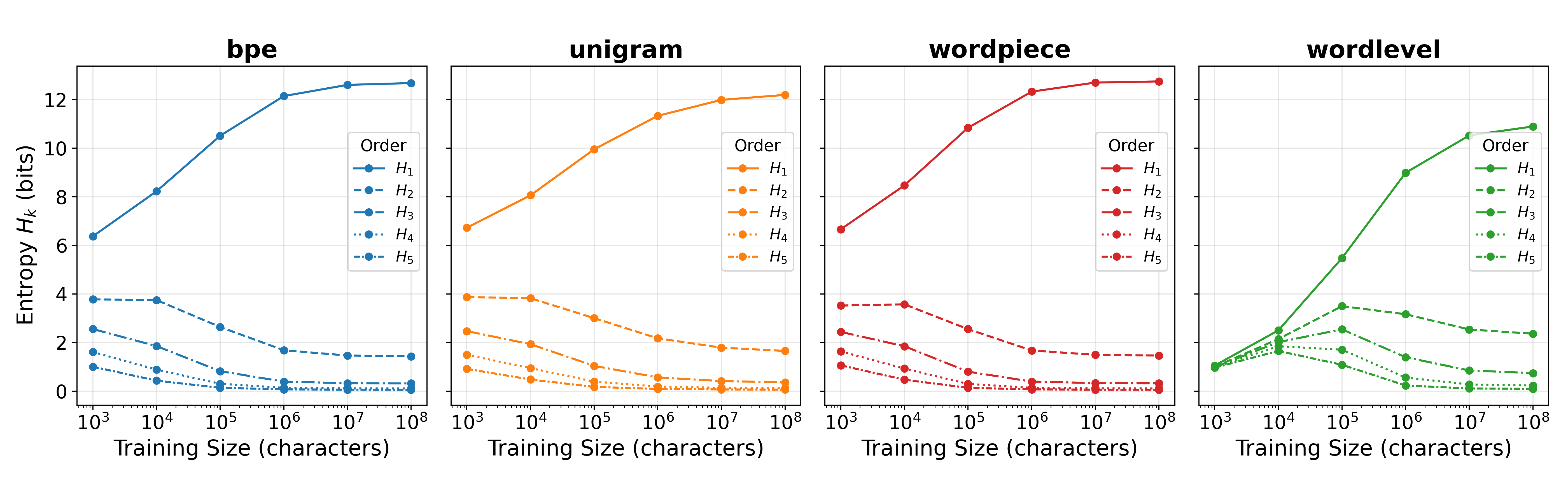}
    \caption{Tokenizer $k$-gram entropy results trained and tested on Turkish for vocabulary size 64k. \looseness=-2}
\label{fig:ent4}
\end{figure*}

\begin{figure*}[ht]
\centering
    \includegraphics[width=0.75\linewidth]{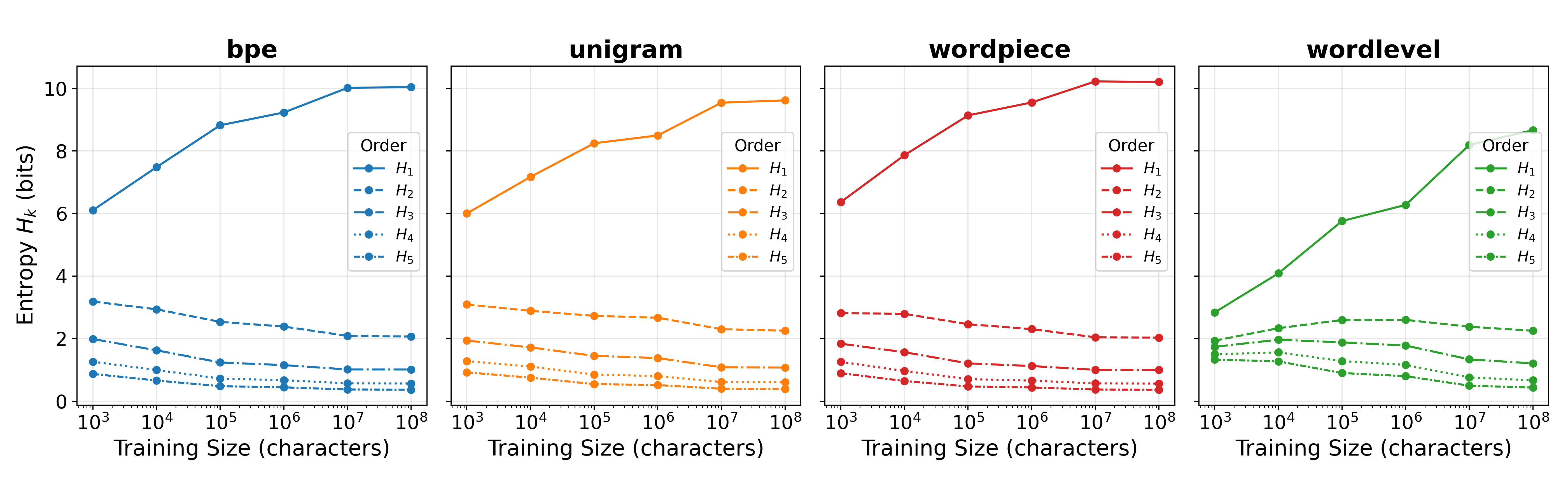}
    \caption{Tokenizer $k$-gram entropy results trained and tested on Code for vocabulary size 64k. \looseness=-2}
\label{fig:ent5}
\end{figure*}

\begin{figure*}[ht]
\centering
    \includegraphics[width=0.75\linewidth]{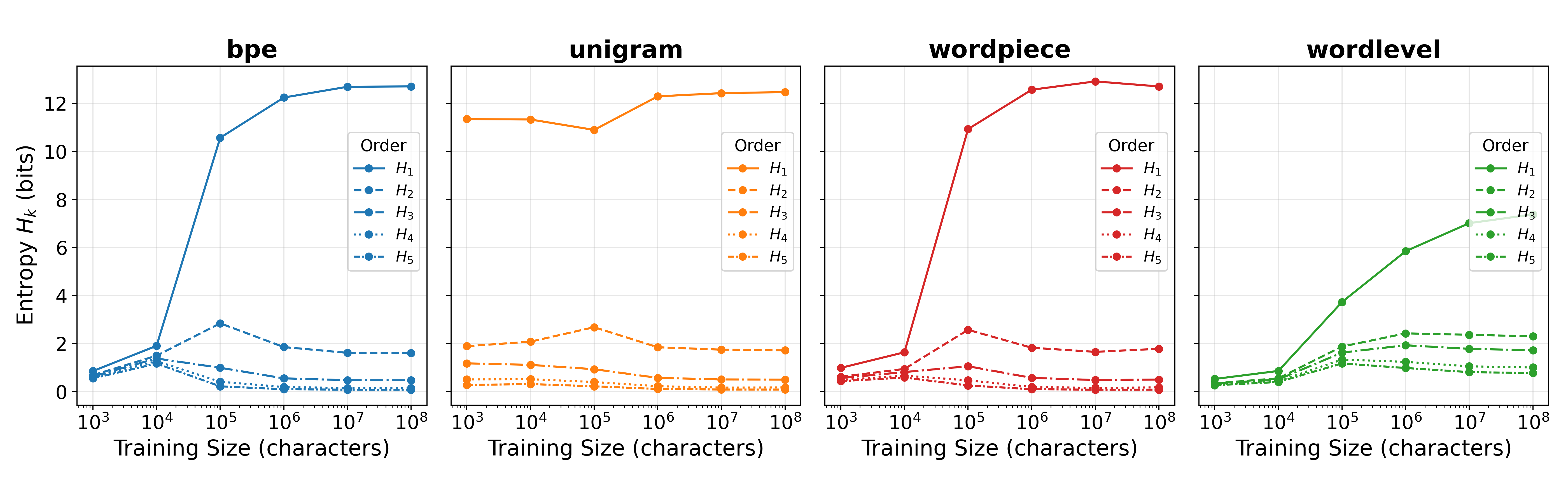}
    \caption{Tokenizer $k$-gram entropy results trained and tested on Chinese for vocabulary size 64k. \looseness=-2}
\label{fig:ent5x}
\end{figure*}

\begin{figure*}[ht]
\centering
    \includegraphics[width=0.75\linewidth]{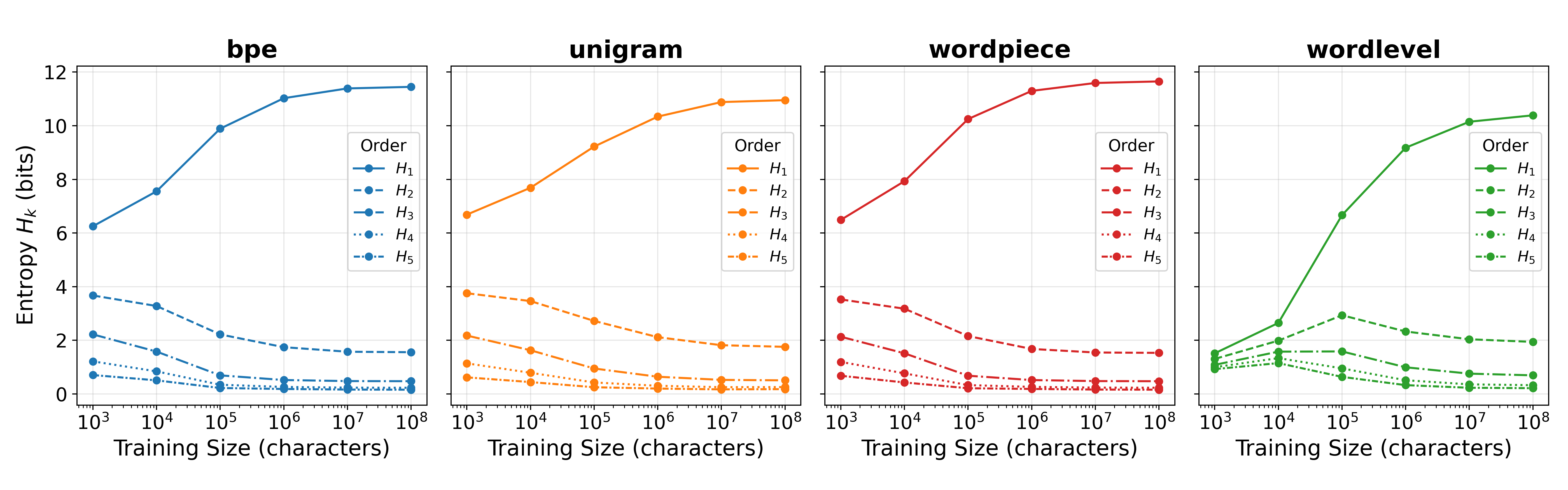}
    \caption{Tokenizer $k$-gram entropy results trained and tested on Chinese-Latin for vocabulary size 64k. \looseness=-2}
\label{fig:ent5y}
\end{figure*}

\begin{figure*}[ht]
\centering
    \includegraphics[width=0.75\linewidth]{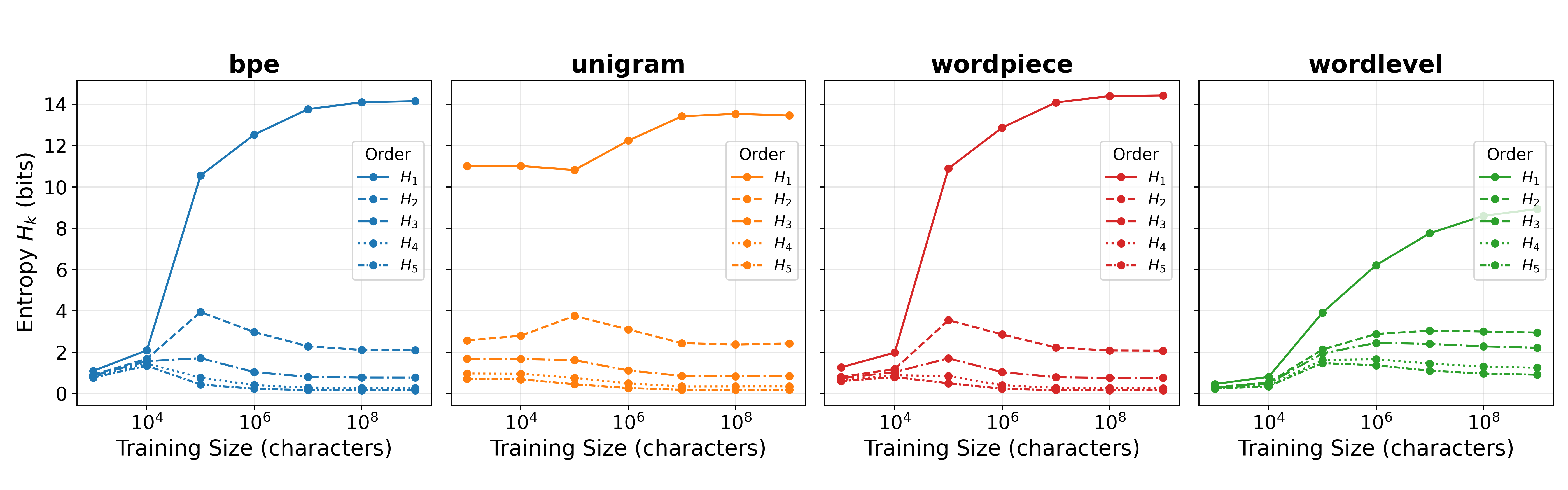}
    \caption{Tokenizer $k$-gram entropy results trained and tested on Chinese for vocabulary size 500k. \looseness=-2}
\label{fig:ent5z}
\end{figure*}

\begin{figure*}[ht]
\centering
    \includegraphics[width=0.75\linewidth]{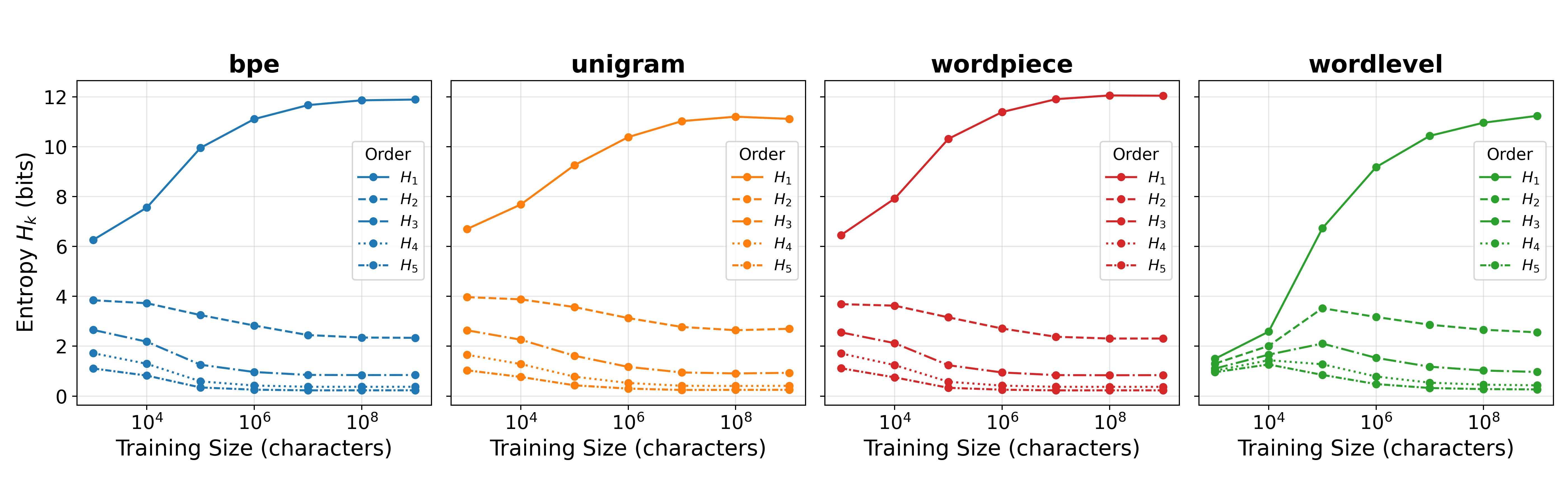}
    \caption{Tokenizer $k$-gram entropy results trained and tested on Chinese-Latin for vocabulary size 500k. \looseness=-2}
\label{fig:ent5t}
\end{figure*}

\begin{figure*}[ht]
\centering
    \includegraphics[width=0.75\linewidth]{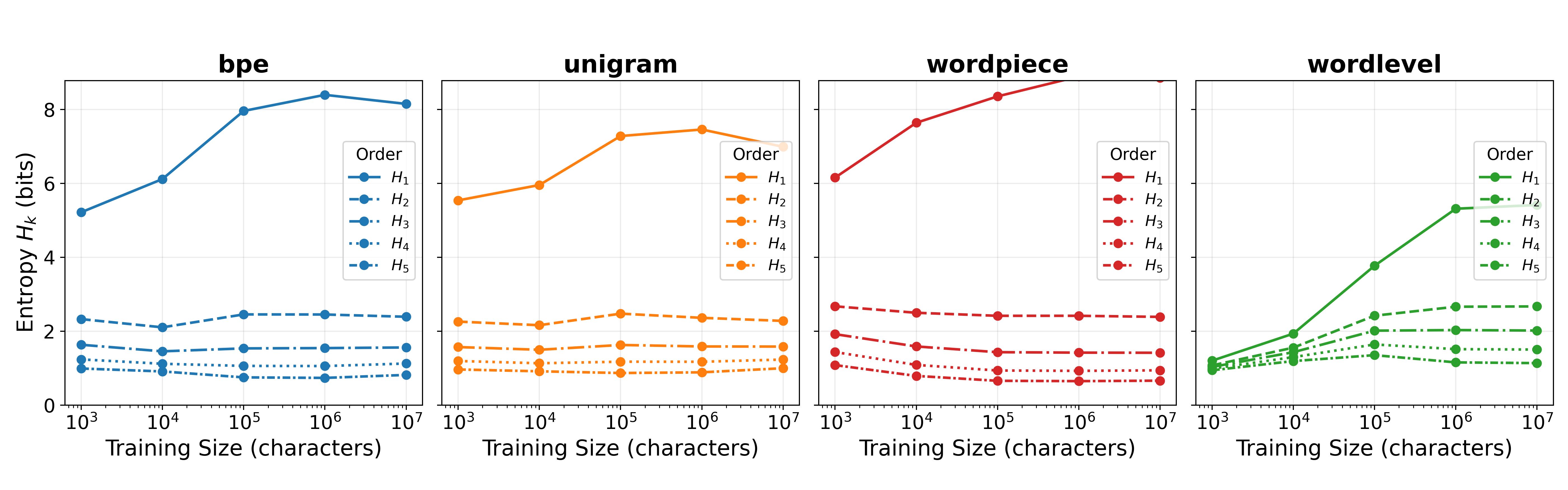}
    \caption{Tokenizer $k$-gram entropy results in domain mismatch for vocabulary size 16k. \looseness=-2}
\label{fig:distribution_turkish}
\end{figure*}

\begin{figure*}[ht]
\centering
    \includegraphics[width=0.75\linewidth]{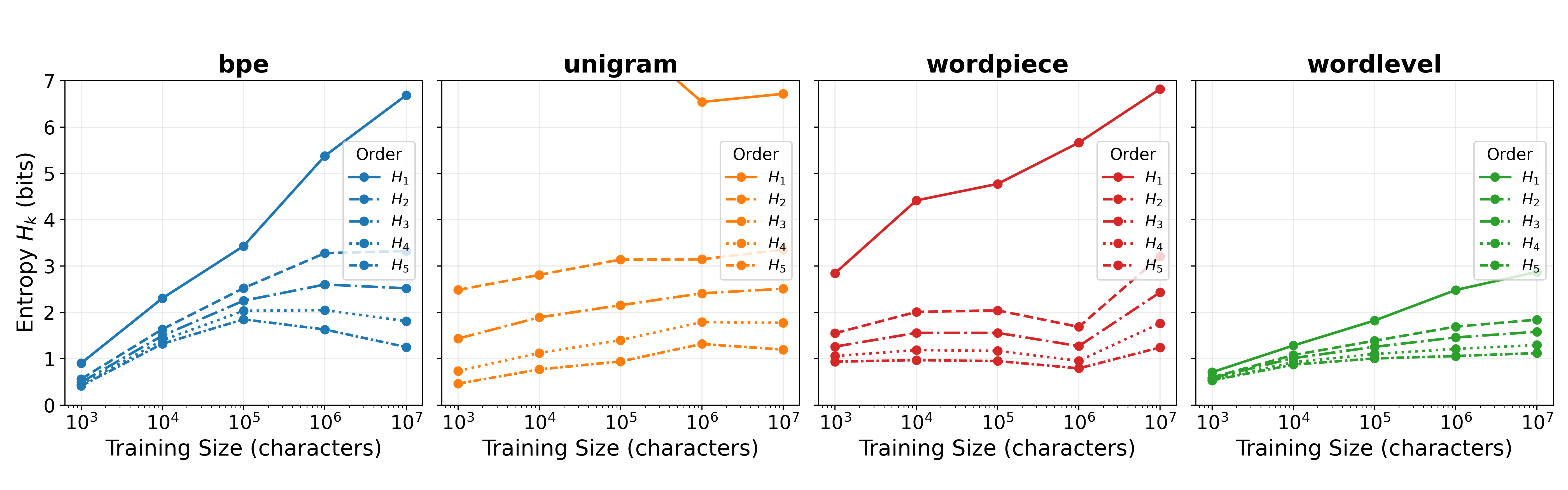}
    \caption{Tokenizer $k$-gram entropy results in domain mismatch for vocabulary size 16k. \looseness=-2}
\label{fig:distribution_turkish}
\end{figure*}

\begin{figure*}[ht]
\centering
    \includegraphics[width=0.75\linewidth]{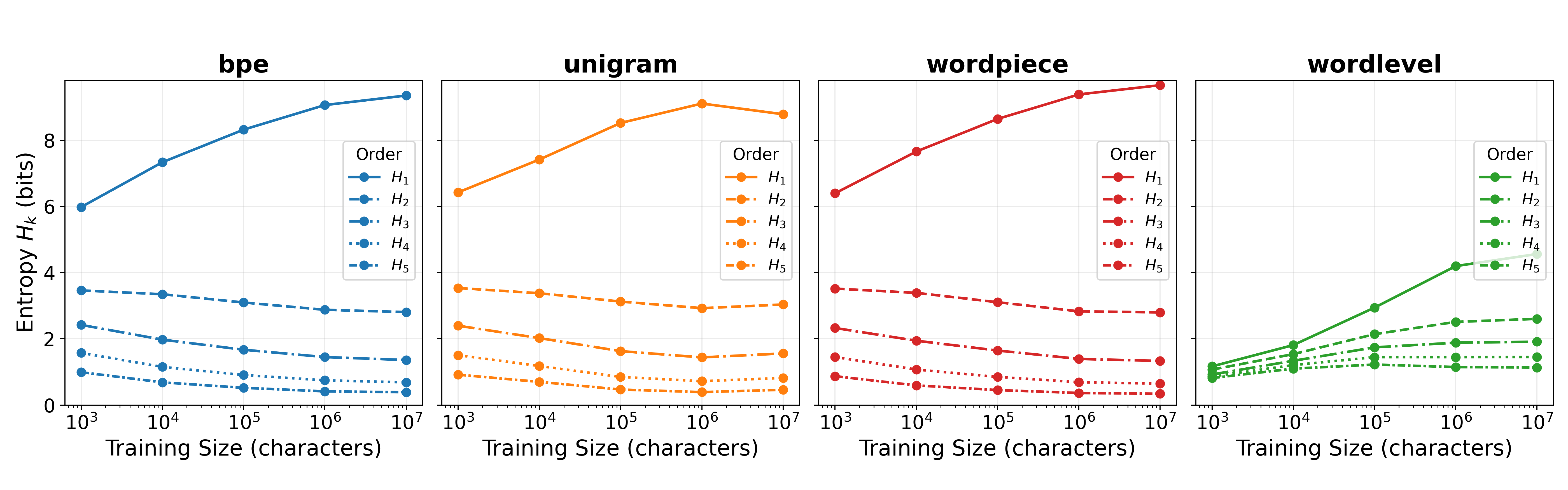}

    \caption{Tokenizer $k$-gram entropy results in domain mismatch for vocabulary size 16k. \looseness=-2}
\label{fig:distribution_turkish}
\end{figure*}

\begin{figure*}[ht]
\centering
    \includegraphics[width=0.8\linewidth]{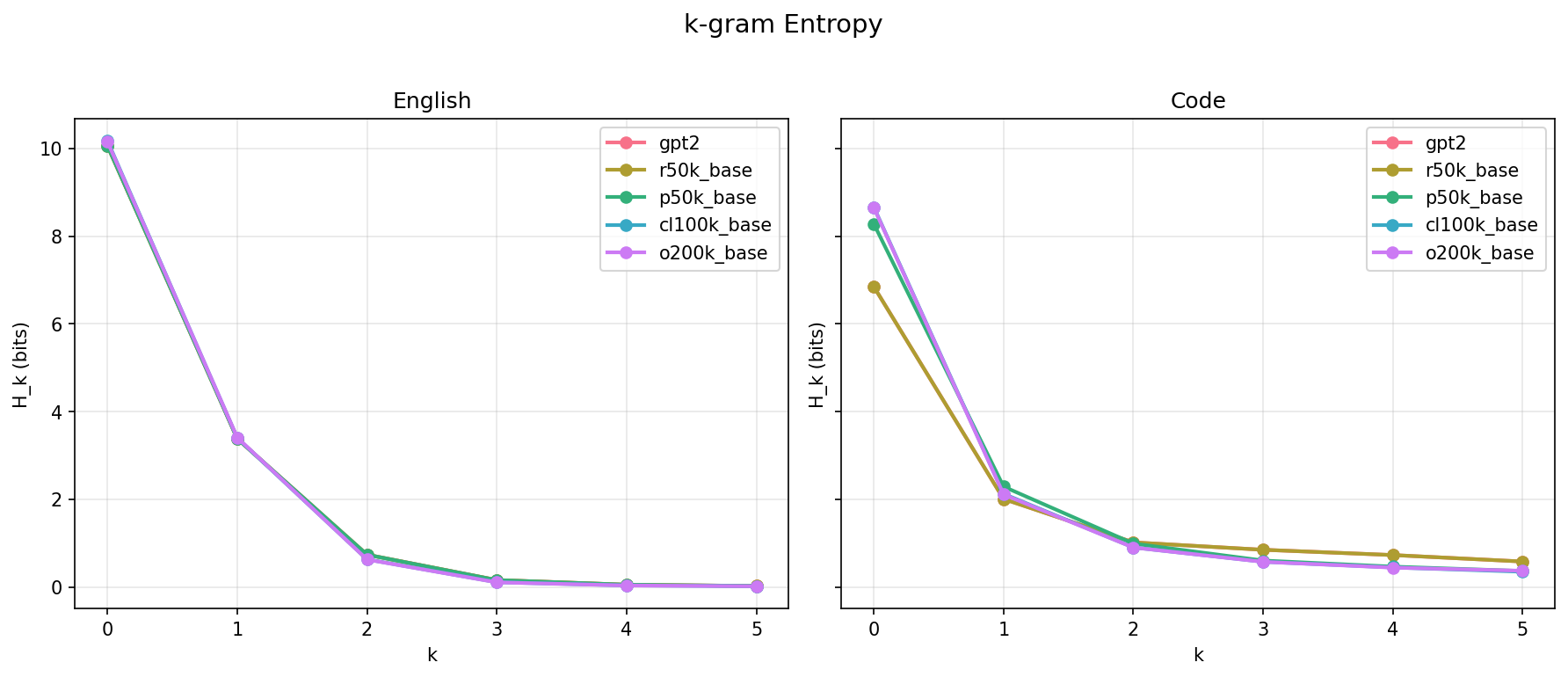}
    \caption{$k$-gram entropies for pre-trained GPT tokenizers. \looseness=-2}
\label{fig:gpt-tok}
\end{figure*}

\end{document}